# New corner stones in dissipative granular gases

## On some theoretical implication of Liouville's Equation in the physics of loose granular dissipative gases

## P. Evesque

**Lab MSSMat, UMR 8579 CNRS, Ecole Centrale Paris**
**92295 CHATENAY-MALABRY, France,** e-mail: evesque@mssmat.ecp.fr

### Abstract:

*The dynamics of a granular dissipative gas is discussed starting from the Liouville's equation to derive a Boltzmann's equation, taking account of the inelasticity of collisions with the walls and between balls . It is recalled that the Boltzmann's distribution, i.e.* exp[-mv²/(2kT)]*, is a steady solution of Boltzmann's equation only when collisions are elastic; hence it is not applicable in the case of dissipative granular gas.*

*Then experiments on non interacting balls in a vibrated cylindrical box are re-examined using cells containing 1 ball or 2 balls. They allow to study the effect on the dynamics of the dissipation during ball-wall collisions. In a first experiment with an electromagnetic vibrator on earth or in board of Airbus A300 –0g of CNES, the 1-ball dynamics exhibit little transverse motion and an intermittent quasi periodic motion along the direction parallel to the vibration. It is quite different from the erratic motion predicted for the Fermi case, with no dissipation. The reported behaviour proves a significant reduction of the phase space dimension of this billiard-like system from 11-d to 3-d or 1-d. It is caused by dissipation, which generates non ergodic dynamics. It exemplifies the coupling between translation and rotation degrees of freedom during the collisions with the walls, due to solid friction at contacts. This eliminates ball rotation and freezes transverse velocity fluctuations. This trends is confirmed by 3-d simulations with JJ Moreau discrete element code, and by a two-ball experiment performed under zero-g conditions in the Maxus 5 flight. For this second experiment, the quasi-periodicity is found much greater, which is probably due to an improvement of experimental conditions. The two balls are not in perfect synchronisation showing the effect of small random noise; but the two particles have never collided. This is then the normal dynamics of a gas of non-interacting dilute spherical grains in a vibrated rectangular container.*

*The dynamics of interacting particles with dissipation is then studied experimentally in the case of a small number of grains, i.e. of small ball-ball interaction, in a cell with a vibrating piston. The interpretation is re-examined and modified. The typical speed of a ball is found to vary linearly with the piston speed* bω*, but decreases when the number of balls* N *is increased (N=12, 24, 36 or 48). The distribution of waiting times* τ *between ball-gauge collisions is found to follow an exponential distribution experimentally, i.e.* p(τ)∝ exp(-pₒτ)*, proving the uncorrelated motion of balls. The amplitude* I *of the ball-gauge impacts has been determined from the signal response of the sensor. This requires to determine a transfer function and to proceed to a deconvolution. The N=12 balls case is used for this purpose. The distribution* f(v) *of ball speed* v *exhibits an exponential trend f(v)=exp(-v/vₒ) in the case N=24, 36, 48. This is temptatively explained using a model "à la Boltzmann" associated with the notion of "velostat". Also a second model is proposed, which describes the fast speed tail of the distribution which is determined to leading order. It is found experimentally that both,* vₒ *and* pₒ*, depend on N, and a scaling law is proposed, although the scaling is tested in a very small range 12<N<48 .*

*At last a general discussion is tempted in the framework of Boltzmann's equation.*

*It turns out that coupling between rotation and translation cannot be neglected in the collisions, because it generates efficient dissipation; that makes the system quite sensitive to even a small number of collisions. This shall be introduce in simulations codes.*

# 1. Introduction

Recently, a tremendous amount of works has been performed to investigate properties of granular gas (see [1] for a review). This problem is a very fundamental one because it is a frontier of statistical mechanics when dissipation becomes dominant; in other words it asks the problem of what dissipation changes in the behavior of a statistical ensemble of particles in interaction: within which limits can one use the analogue of thermodynamics quantities to describe the behavior of these systems?

Indeed, dissipation is already taken into account in many different problems of physics, such as electronic transport, or as the hydrodynamics of Newtonian fluids; analogy with these problems has been the idea of the theoretical understanding of granular gas [1-3]. So, one starts the description of granular gas with a series of continuity equation imposing preservation rules of mass, momentum and transfer of energy, and assumes a distribution of speed for the particles, which is usually taken to be almost Maxwellian because the systems are assumed in a local thermodynamics equilibrium.

However, the dissipation at work in these granular systems is different from the other ones (Newtonian fluid, electronic transport,…) because dissipation intervenes directly in the collisions. That cancels probably the validity of a series of approximations as shown later, and despite what is often supposed [1].

Indeed, many results demonstrate the atypical behaviour of granular gases: For instance, dissipation generates clustering in granular gas; but this effect occurs at very small density of grains as shown by a recent micro-gravity experiments [4] or using a pseudo-Maxwell demon experiment on earth [5]; the interpretation given [6] links this clustering to the change from a Knudsen regime (for which particles interact mainly with the boundaries) to the case of interacting particles. The photos of granular gas in 0g [4, 6] show also that the wall speed is faster than the typical ball speed even when the density is so small that no cluster occurs, so that the excitation is of "supersonic" nature [6]. This asks the validity of a continuous description; and it asks also about the true role plaid by boundaries: do they act as a thermostat or as a "velostat" [7], *i.e.* do they impose a speed or a kinetic energy? Also, do the system behaves in a similar way on earth as in micro-gravity?

In other words, many questions remain to be answered, and they shall be asked from the different points of view theory, simulations and experiments:

• For instance, from a theoretical point of view: can one use without adaptation thermodynamics concepts and standard results, such as the temperature and the Maxwell distribution… to describe granular gas? Can one use also continuous modeling? Does gravity play no part? What are the boundary conditions?

• From a numerical point of view: are all the microscopic models of collision equivalent? Do they predict the same behavior? Can one discriminate between them? How do they comfort theoretical understanding? How well do the simulations predict experimental results? Are 1-d, 2-d or 3-d numerical simulations equivalent? Is earth gravity important? Does the discrepancy between experiment and simulation come from ill controlled experiment or from a lack of the simulation technique, from wrong collision rules…? Most simulations neglect rotations; are they right?

• From an experimental point of view: are the experiments well controlled? Does one measure truly what one expects?

And the final question is: how to use all these results to converge towards a correct understanding of the physics of a dissipative gas. But, obviously, a first step of testing has to be done with a little number of grains to gain on accuracy of criteria and discriminate between hypotheses, because statistics and randomness smoothen and mix the differences.

This paper proposes a frame work to discuss some basic assumptions on the theory of dissipative granular gas, and shows with a simple experiment that a deep understanding about the limitations can be gained from very simple experiments (or simulations); this necessitates to avoid the use of unproved analogies. The paper focuses on 1-ball and few-balls experiments. The 1-ball experiment demonstrates that rotation cannot be forgotten in the modeling of collisions. The other experiments deal with a little number of grains N<50, whose statistical behaviour is described, since their simulations should be considered as natural benchmarks.

The paper is built as follows: in section 2 some example about the lack of knowledge on the behaviour of granular gas is recalled. Section 3 describes the formalism of Boltzmann's equation applied to very dissipative systems. Section 4 describes and discusses experimental results in the limit of an infinite dilution, which corresponds also to the 1-ball case; some of these results can be found on [8,9,10]. Section 5 describes experimental results on the dynamics of few grains in a vibrated container [11]; new interpretations are given. Section 6 addresses the question of predicting the density probability function of the 1-particle in the case of few grains.

## 2. Few lacks in the literature on dissipative granular gases

Despite the tremendous amount of published works in the domain of granular gas, there is still little amount of complete answer one can gain from the literature at the moment (end of 2003). For instance if one is interested by the problem of simulations, one can get from the literature different procedures to simulate the gas of particles and its dissipation, starting with different collisions rules; one uses (i) simply a normal restitution coefficient, or (ii) two coefficients (a normal and a tangent restitution coefficient), while others (iii) include particle rotation, solid friction, …. But none of these works tend to define clearly the minimum ingredients to get the true limit behaviour of a dissipative granular gas as it is encountered in experiments.

None also tries and solves numerically the complete real problem with as little approximation as possible, to establish the correct "standard" behaviour and its dependence on different parameters. Indeed, nobody can assert that experiments are under complete control as can be a numerical simulation; also the experiments cannot give access and measure all the statistical quantities as it can be done in numerical simulations. So, this should have been the priority in the very starting procedure to use simulations to allow (i) finding efficient relevant macroscopic quantities to characterise the physics obtained, (ii) defining the process of a correct comparison with experimental results, (iii) checking the quality of the experiments, (iv) finding the very fundamental ingredients which have to be included to keep constant the physics.

This is also the only way to define progressively simplified models with their own domain of applications. But, it is not available at the moment in the literature, where one finds often papers which mix experimental results and simulations without measuring the experimental sets of parameters in order to inject them in the simulations …. Comparing experiments and simulations is not aimed at noticing punctually some effect or artefact, but when noted to carefully study the effect or artefact to be sure that it does not hide or involve some more important point and other phenomena.

This is true for collision rules as already told; but there are other examples: (i) it is considered in general that the moving box plays the role of a thermostat, but it may be better described as a velostat, because it imposes likely the bead speed [8] instead of the bead energy. (ii) What about lateral walls? (iii) Gravity has to play an important role also, whose effect on granular gas behaviour should gain to be clarified; indeed results in 1-g and in 0-g are often compared, but the timescale and confinement are strongly affected by gravity; for instance the time scale of a bouncing varies as $\tau = v/g$ in 1-g, while it scales as $L/v$ in 0-g, which changes the sensitivity of the high speed dynamics of the system to a fluctuation of the period of excitation…; and the particles are confined on the bottom in 1-g while confinement requires two opposite walls in 0-g. This comes back to the problem of the boundary conditions; as shown in section 4, the presence of a second walls is crucial [8-9],…. Indeed no complete understanding exists still.

## 3. Basis for the dynamics of dissipative granular matter in a vibrating container

In general one establishes a strong parallel between the dynamics of a granular gas and the thermodynamics of a classic gas. One starts from macroscopic conservation rules as those obtained from hydrodynamics. One introduces also a local distribution f(r,p,t) of speed in order to take into account the microscopic issue; however, owing to the lack of knowledge about this distribution in the present case, it is most often taken as Maxwellian, assuming that a granular gas behaves approximately as a classic gas or liquid, despite the dissipative nature of the collisions.

In this section, I want to demonstrate the strong difference between a dissipative system and a non dissipative one. So the problem is settled in a reverse way, starting from a correct statistical mechanics description of the motion. So, I start with the Liouville's equation of motion, which is the correct procedure; then I establish some kind of Boltzmann's equation from the Liouville's one, introducing collisions. Solving this equation with stationary conditions shall lead to the correct distribution function. However it is shown that this distribution can not be Maxwellian, because of the dissipative nature of the collisions.

## 3.1. The Boltzmann's equation

The mechanics of granular materials obey the general laws of mechanics; here it can be described as free flights of particles in between collisions. When the time spent by a particle in free flight is much larger than the duration of the collisions, the number of complex collisions engaging 3 or more bodies at the same time are very rare so that the collisions can be considered as binary collisions. In this case one speaks of the granular gas mechanics. However, the main difference with classic gas is the dissipative nature of these collisions. Hence it is quite important to estimate properly the collisions rules and to characterise and quantify the effect of the dissipation. An other characteristic of granular gas is the local nature of the interactions, since two balls interacts only by contact (if one except some electric charge effect, or some effect generated by some interstitial fluid). For instance, be v and m, the speed and mass of a typical ball; be $\tau_{coll}$ the duration of a typical collision and $l_c$ the mean free path between two collisions; as elasticity theory tells that $\tau_c$ scales as $\tau_{coll} \approx m^{2/5} v^{-1/5}$ and since the time $t_{coll\text{-}coll}$ in between two collisions scales as $l_c/v$, one gets $\tau_{coll} / t_{coll\text{-}coll} \approx m^{2/5} v^{4/5} / l_c$ that tends to 0 with v; hence one deduces that the limit of the dilute gas shall be obtained at small speed, as soon as capillarity effect and electric charge effect do not happen.

Within this approximation of "dilute" condition, the solution of the N-particle problem can be integrated over the N-1 or N-2 particles which do not interact with the considered particle [12,13]; hence the problem can be reduced to the one of finding the distribution function at time t of a single particle f(q,p,t) and its evolution; here q and p are the position and momentums of the particle, in which rotations are included. This simplification requires also that topological constraint between successive collisions are small enough to allow neglecting correlations in between successive collisions.

Within these approximations, and starting from the complete Liouville's equation [12,13] which describes the evolution of the complete distribution function with N particles, one can show (see Appendix A) that the evolution of the system can be described by the distribution function f of a single particle, and that f is described by the Boltzmann's equation:

$$[\partial/\partial t + \mathbf{v}.\nabla_r + \mathbf{F}.\nabla_p] \, f(p,q,t) = (\partial f/\partial t)_{coll} \qquad (1)$$

Here the term $(\partial f/\partial t)_{coll}$ contains collisions between two balls and between the balls and the container wall, since both collisions can happen. As the container is vibrating, its dynamics can be described as the oscillation of a spring-mass system with Hamilton equation $H = p_w^2/(2M_w) + kq_w^2/2$; the ratio $k/M_w$ and the energy $E_o$ shall be adjusted to get the right frequency and amplitude; it is supposed here that $M_w \gg Nm$, m being the particle mass, to neglect the variation of amplitude due to repeated collisions.

As mentioned above, Boltzmann's equation is derived directly from the complete Liouville's equation of motion after averaging over the N-1 other particles ; hence it is exact within its approximations. When the density of particles gets larger and that three-body, or 4-body,... collisions become more numerous, one can use the BBGKY expansion of the Liouville's equation to compute the evolution and to incorporate cross correlations between the motions of particles.

Nevertheless, the problem is completely solved in the dilute regime when one finds the correct distribution function f. It is necessary to write it down correctly. It takes the form:

$$(\partial f/\partial t)_{coll} = \int dp_1 \int d\Omega \sigma(\Omega) \, [|v'-v_1| \, f(p')g(p_1)- |v-v_1| \, f(p)g(p_1)] \qquad (2)$$

The right term of Eq. (2) is made of a sink term and of a source term; the sink describes collisions between both particles p and $p_1$ (leading to a rate of disappearance of p) and the source term describes collisions between particles p' and $p_1$ generating a particle with momentum p. $\sigma(\Omega)$ is the cross section of the collisions. The use of g(p) in Eq. (2) allows to treat at the same time collisions with the wall, g≠f, and collisions between balls, g=f.

## 3.2. *Classic gas of atoms:*

At this stage it is worth recalling some classic results about the statistical mechanics of a gas of atoms. Indeed, in the case of a gas, one looks for a stationary homogeneous solution of the Boltzmann's equation, *i.e.* $\partial f/\partial t=0$, $\nabla f=0$. This imposes to search f satisfying $(\partial f/\partial t)_{coll} =0$ . But collisions between atoms obeys some simple rules of conservation, which are mass preservation, momentum preservation and energy preservation; furthermore, one gets also $\sigma(p_1,p_2|p'_1,p'_2)= \sigma(-p'_1,-p'_2|-p_1,-p_2)= \sigma(-p_1,-p_2|-p'_1,-p'_2)= \sigma(p'_1,p'_2|p_1,p_2)$ . Hence Eq. (2) can be transformed into:

$$(\partial f/\partial t)_{coll} = \int dp_1 \int d\Omega \sigma(\Omega) \, |v-v_1| \, [f(p')f(p'_1)- f(p)f(p_1)]=0 \qquad (3)$$

So, looking for a stationary homogeneous solution of Eq. (1) leads to look for a solution of Eq. (3); Eq. (3) is satisfied in turn if $f(p')f(p'_1)= f(p)f(p_1)$. This last condition is satisfied if f is such as its logarithm log(f) is a linear combination of the invariant of the collisions. Hence one finds the general solution log(f)= Am + Bp +Cp²/2m, with A, B, C being 3 constant; this condition can be rewritten in the form of the Maxwellian distribution:

$$\ln(f)=D-(p-p_o)^2/(2mk_BT) \qquad (4.a)$$

$$\text{or} \qquad f = A_o \exp\{-(p-p_o)^2/(2mk_BT)\} \qquad (4.b)$$

Eq. (4) is then a possible stationary solution. It remains to demonstrate that Eq. (4) is the solution.

This is straight forward applying the H theorem of Boltzmann [12,13]. This H theorem tells some restriction on the evolution of any function f obeying the Boltzmann's equation with collisions obeying the conservation rules quoted above, *i.e.* preservations of mass, momentum and energy: the H theorem tells that the local entropy $\int -f \log(f)dv$ shall increase continuously; so, as Eq. (3) describes a stationary state, Eq. (4.b) is the exact solution.

### 3.3.  Extension to the case of little dissipative systems:

When the system dissipates little and the number of collisions is large, the H theorem applies at least approximately at the local scale leading to a local convergence time which is short compared to the complete evolution time. In this case, the procedure to find f consists to apply a perturbation approach starting with a Maxwellian unperturbed distribution. This is often used in the case of electrons in metal, of the hydrodynamics of Newtonian fluids… [13]

The reason why this approximation is valid comes from the great number of collisions which preserve energy; hence the system converges locally fast towards its local equilibrium; then transport takes place. And the dissipation occurs due to the slow diffusion of heat.

### 3.4.  The case of very dissipative systems is likely different:

However, when each collisions dissipate energy, the system dissipates efficiently, and the problem changes likely of nature because the collision rules do not obey anymore the law of conservation of energy. The first consequence is probably that the equilibrium distribution f is no more Maxwellian, because this distribution requires that energy is preserved during collisions, *cf.* demonstration of Eq. (4).  An example will be given when looking at the problem of few grains in the second next section, *i.e.* §-5.

An other well-known effect caused by dissipation is the non-preservation of the elementary volume in the phase space. This comes directly from the Liouville's equation and the Hamilton equation of motion. That leads to concentrate the dynamics on attractors, which are periodic orbits, strange attractors… Indeed many typical examples are given in textbook with simple oscillators, leading to circular or tore orbits, and chaotic dynamics...

Here it is proposed a simple example of the reduction of the phase space; it has been obtained when studying the dynamics of a single ball in a vibrating box to calibrate a force sensor. It can be seen also as the limit case of a granular gas with particles interacting only with the walls.

## 4. The case of a single ball in a container with a vibrating piston.

The limit of a gas of non interacting particles can be studied when diminishing the density of the system, till the particles never meet. An other way is to reduce the dimension of the cell and to work with a very little number of particles. This is what is reported here. In this case one expects the system is connected to the problem of chaotic billiard [14]. In 1d with no dissipation and in 0g, it becomes also the Fermi problem [15-17]. Or it is the bouncing ball case, when the sytem is 1d under gravity and no top wall [18,19]. This 1-ball example in a vibrating box can also be used to test the effect of dissipation at boundary, in order to define correct boundary conditions. Is the box playing the role of a velostat and imposes a definite mean speed [7], or is it a thermostat…?

However, this experiment has been performed at first in order to calibrate a gauge, to measure accurately the speed of the particles from a series of impacts and to determine the restitution coefficient $v_{out}/v_{in}=-\varepsilon$ as a function of these speeds. This has been achieved, but it turns out that it leads to new understanding: It demonstrates also that rotations generate strong dissipation during collisions that affects the whole mechanics of granular gas.

The efficiency of the calibration is linked to the fact that one gets a pure 1-d dynamics, with little transverse motion. Hence the impact times correspond to well determined travel lengths and speeds.

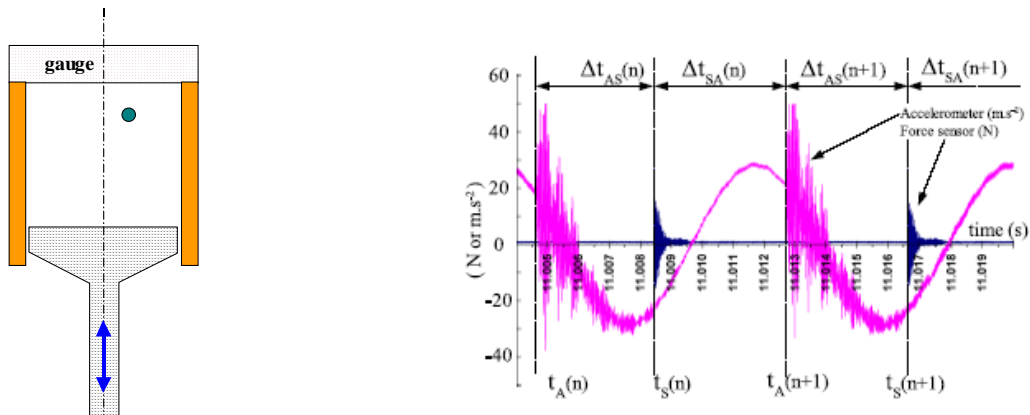

**Figure 1:** The experimental set-up: a sensor is placed on top of a cylindrical fix cell (H=10mm, D=13mm); this cell contains a stainless steel sphere (d=2.0 ±0.002mm); it is closed on top by a force sensor and on bottom by a vibrating piston on which an accelerometer is fixed. This allows measuring the impact times of the ball with the piston and with the lid, (from [9]).

### 4.1. Experimental technique and typical result.

As shown in Fig. 1, a PMMA-made transparent cylindrical cell of length L=10mm, diameter 2R=13 mm, is fix in the lab frame; it contains a sphere (radius r=2.0mm); its bottom is a vibrating piston ($z=b \cos(\omega t)$, 30Hz<f=$\omega/(2\pi)$<120Hz) driven by a computer-controlled electro-magnetic vibrator (Model V450 from LDS with its 1kW amplifier LDS, Model PA1000L) and it is closed on top by an ICP sensor  (Model 200B02 from PCB Piezotronics), to be calibrated. A triaxial accelerometer (Model M356A08from PCB Piezotronics) is fixed to the piston (from AISI 316L stainless steel, 12.7mm diameter), that measures $b\omega^2/g$ and the frequency $\omega/(2\pi)$ of the movement. Its signal is perturbed by the impact of the ball with the piston, so that it allows measuring these times of impact $t_{2n}$ and the position $z_{2n}$ of the piston. Data are acquired with 12bits at 2MS/s acquisition rate, using the simultaneous-sampling multifunction DAQ device (National Instruments, Model PCI-6110) and Labview.

The impact sensor is fixed in the lab frame; this limits the perturbation of its response due to inertial effect. It allows then determining the times $t_{2n+1}$ and the amplitude $I_{2n+1}$ ($=I_n$) of impacts with the sensor. Examples of these signal are given in right part of Fig. 1. One observes a rather complex structure of the signal generated by each collision, with beats of 100kHz and 110 kHz modes. Anyway, each impact starts with a short pulse of amplitude I constituted by rapid increase and decrease, which lasts the time $\delta t \approx 6\pm 0.5\mu s$. This peak is followed by large oscillations at 100kHz and 110 kHz which beat and which decay with a characteristic time of $\approx 0.1$ms.

A video camera with a resolution 700*500 pixels has been used associated to stroboscopic illumination to observe the bead motion in a central zone (14 mm heigth) of the glass cylinder and the moving piston. When the cell size is minimum the distance between the mean position of the piston and the force sensor is 10 mm and the total cell is in the field of the video camera. In addition, stereo viewing analysis can be performed if needed by using a Hi-8 camescope with a viewing angle different from the preceding one.

## 4.2. Experimental collision restitution coefficient ε.

One can then compute $v_{n+}$ and $v_{n-}$, before and after each impact n, and determine $\varepsilon_n$=-$v_{n+}/v_{n-}$ as a function of $v_{n-}$, for the gauge and for the piston. A typical example is given in Fig. 2 in the case of the sensor. Similar results are obtained with the piston.

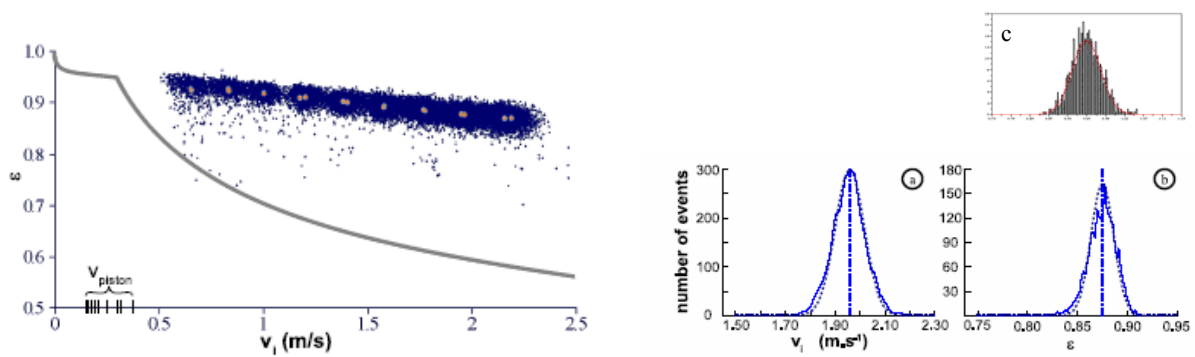

**Figure 2: Left:** Variation of the restitution coefficient ε as a function of the ball speed, for a 2mm diameter stainless steel sphere (d=2.0 ±0.002mm) hitting the fix sensor.
**Right:** Statistical distributions of the input velocity (a) of the ball with the sensor and of the restitution coefficient ε (b) . The dotted curve corresponds to the Gaussian profile of same half-width at half-height. On top of (a) and (b), fig (c): normal dispersion on ε obtained from 1-d simulation when adding some Gaussian noise on the measured time (the accuracy is $10^{-4}$ s on the piston and $10^{-5}$s for the sensor). (from [9] except for fig. c which is from Cl. Ratier, G. Thoorens)

The noise observed on ε (Fig. 2.b) can be due to some uncertainty on the measurement of the impact times, as demonstrated by some simulations reported in Fig. 2.c. However, similar dispersion can be also generated by some random fluctuation of the mean position and of the mean amplitude of the piston, or can be due to g-jitter in micro-gravity condition. The lack of sphericity (δd=2μm, d=2mm) of the ball or the surface roughness of the piston can also generate some dispersion by introducing some rotation momentums during collisions, which may increase dissipation. As shown on Fig. 2c, similar normal dispersion curves on ε have been obtained from 1-d simulation when adding some Gaussian noise $δt_1$ and $δt_2$ on the measured time ($δt_1=10^{-4}$ s for the piston and $δt_2=10^{-5}$s for the sensor which are approximately the resolution from the technical data-sheet of the manufacturers).

Under the experimental condition, one observes a periodic motion with 1 roundtrip per period. Indeed a periodic motion occurs when b/L is large enough) *i.e.* b/L> $α_{thres}$) as can be obtained from 1-d simulations [20], or from stability analysis of the periodic condition as it will be told later. So, this has been observed experimentally too, with some resonant rate that depends on the quality of micro-gravity condition and/or of the device (a device with a crank and a connecting rod has been used in the Maxus 5 experiment, that leads to less fluctuations). Theoretical description of stability of the 1-d periodic motion is reported in Appendix C.

- *Effect of collisions on rotation and transverse motion:*

But the main result is not there: one observes also the freezing of the transverse motion of the ball as soon as the ball hits the lid. Indeed, this can not be demonstrated from the electronic signals of the sensors, but has to be observed optically either with the eyes or with a video camera: performing the same experiment on earth at small enough amplitude, or without the lid, one sees the ball jumping from left to right erratically; but as soon as the ball starts hitting the wall the vertical trajectory is stabilised and it evolves no more transversally (one still observes a very slowly lateral jittering, *i.e.* with a random transverse speed $v_⊥$ of amplitude $v_⊥$ < 1mm/s for 100Hz excitation frequency about). This demonstrates the freezing of the transverse degrees of freedom. I will show now that it demonstrates also the freezing of the rotation degrees of freedom.

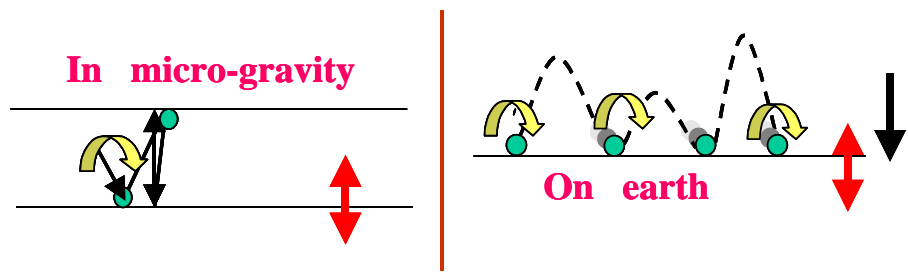

**Figure 3:** Sketch of collisions with rotation with bottom and top walls (left), with only top wall (right)

The stabilisation of the vertical trajectory comes from a coupling between transverse motion and rotation during the collisions. This coupling is due to the solid friction which exists between the surface and the ball and that acts on the ball to tend to impose non sliding condition at the contact point during collision. Let us for instance consider the case without a lid, sketched on the right part of Fig. 3. A transverse motion will impose some rotation after the first bounce; but the next bounce will preserve the same ratio of lateral speed and rotation, so that the ball can keep on moving in the same direction. With a lid things are different: any transverse component of the ball speed will force the ball to rotate in one direction, say spin +, at one wall due to the collision rules, and in the other one, say spin -, at the opposite wall. This improves much the dissipation; hence it blocks both rotation and transverse motion. This is demonstrated in more details in Appendix B.

This effect can be generalised to ball-ball collisions in 2d or 3-d: due to solid friction, and due to the fact that collision are not frontal, collision rules fix the ball spins at the end of each collision from the impact parameter at the beginning of the collision (the impact parameter describes the orientation of the relative translation speed of the balls and the orientation of the contact surface). Next collisions occur from other directions with different impact parameters so that an important part of spinning energy is lost during each collision. This effect is stronger in 3-d than in 2d because the probability of frontal collision decreases with increasing the space dimension; and it does not exist in 1-d. Hence one understands why dissipation looks quite large in 3-d experiment on granular gas, much larger than in a 1-d case. For instance, assuming spherical grains, and assuming a random distribution of kinetic energy, one expects that collision freezes d-2 degrees of translation-and-rotation over the 2d-1 degrees of freedom at the beginning. Hence one expects approximately a 40% energy-dissipation rate/collision. This may change with non spherical shapes.

While it has been known for long that efficient simulation of quasi-static mechanics of granular material requires to introduce rotation motion of individual grains, it is rather strange that no simulation on granular gas has ever concluded of the necessity of taking account of this effect. Here we demonstrate this is essential to model correctly the physics of this system. In practice, until now (end of 2003), most published simulations on granular gas have just imposed some collision rules (nearly at random?) in order to get a better fit with experimental data; very few introduce rotations, some others introduce a tangent restitution coefficient instead. Some others introduce a single speed-dependent normal restitution coefficient, just to fit the experimental results; an example of such a law is given in Fig. 2, which is far to agree with the measured experimental points, but it has been determined [21] to fit and "explain" our own experimental finding [4]. So, simulations simply do not take correctly care of this phenomenon.

The present effect is then a direct result of space research.

♦ *Simulations using DEM of Moreau:* Anyhow, I have performed 3-d simulations with a Discrete Element code from Moreau [22] whose collision rules allow to take account of restitution coefficient and friction during collision ; the simulations have confirmed the experimental finding when friction is taken into account. In particular the stabilisation of the linear path parallel to vibration direction and normal to a wall is obtained when friction is introduced. At first sight, this trajectory remains stable when changing the gravity orientation of few degrees, although there might perhaps be some lateral drift of the trajectory at long time due to the gravity orientation; but this takes too long to be tested directly. I have observed this effect in an other situation, when using vertical gravity and horizontal vibration, and an initial speed characterised by a rapid horizontal component, larger than the vertical component; in this case a transient horizontal trajectory is generated after few lateral bouncing; this transient trajectory is linear, quasi periodic and oriented perpendicular to the direction of vibration; this transient trajectory falls down slowly. Furthermore some transient trajectories which are linear and quasi periodic and which are oriented perpendicular to the direction of vibration have also been found, for peculiar initial conditions; this reveals the existence of different cyclic trajectories with their own distinct basin of attraction, with a hierarchy of Lyapounov exponents.

♦ ***Effect of the lid : reduction of the phase space dimension:*** The present experiment points out also the effect of boundary conditions: indeed, the behaviour obtained in normal 1-g condition without a lid (and in 3-d) is different from the micro-g condition that imposes to use a lid.

The effect of the lid is then to freeze a lot of degrees of freedom since the dimension of the phase space of this experiment passes from 11-d, (which is the number of freedom degrees for a 3-d sphere with time dependent excitation, since one gets 3 positions, 2 angles, 3 speeds, 2 rotation speeds and the time or the piston position), to 3-d, which correspond to z, $v_z$ and $v_{piston}$. Comparatively, the resonance, which is also observed in these experiments, reduces the phase space dimension much less: indeed, the synchronization links together the ball position z to the time (hence to the piston speed $v_{piston}$); it relates also $v_z$ and the impact times $t_n$ to the length L of the box and to the period of the forcing; this forces the attractor to be 1-d instead of 3-d. It diminishes then the space dimension from 3 to 1, *i.e.* 2 units, while the lid diminishes from 11-d to 3-d, *i.e.* 8 units!!

Reduction of dimensionality is classic in dissipative systems with dissipation; this explains in particular chaotic strange attractors; however, it turns out to be peculiarly efficient in the present case.

● ***Intermittency:***

It is worth noting that 1-d modelling of the 1-g bouncing ball problem has concluded to the stability of the periodic bouncing [18] whatever the periodic excitation and restitution coefficient. But this is not verified by experiments, which find a random Maxwellian-like distribution of speed is obtained [20]. This is likely induced by the existence of the small fluctuations.

Indeed, one expects the 1-g experiment without lid to be much more sensitive to fluctuations, since the time of flight $\tau_{f-1g}$ between successive bounces scales as the initial speed after the last bounce $v_n$ , *i.e.* $\tau_{f-1g} \approx 2v_n/g$, while it shall scale as $\tau_{f-lid} \approx L/v_n$ with a lid. So the 1-g case shall behave in such a way as the faster the ball, the longer the flight and the larger the number of periods in between bounces; considering now a small uncertainty $\delta v/v$ ; it shall result in a complete indeterminacy of the phase of the next collision when $v$ is large enough; hence this predicts that the 1g-no-lid system is quite sensitive to fluctuations of excitation period at large excitation… This process of uncertainty production is not possible in a system with a lid. Does this lead to predict that experiments of bouncing in between two walls shall be more stable than the 1-g experiment? This shows at least that the real mechanics of the motion can obey different rules from the expected 1d-no-lid modelling.

Nevertheless, intermittent resonance has been also observed experimentally with a lid, either in 1-g or in mini-g conditions *i.e.* during parabolic flights in board of the Airbus A-300 of CNES, *cf.* Fig. 4. Intermittency is most likely due to the imperfections of the experimental set-up and/or to g-jitter. Fig. 4 reports the variations with the frequency of (a) the rate of resonance, (b) of the mean position of the piston and (c) of the mean length of the cell. This shows that the larger the frequency (hence the larger the ball speed), the larger the periodicity rate, the less the intermittency, the smaller $<z>/b$ and the larger the piton speed $b\omega \sin(\omega t + \varphi)$ at impact.

These results might be understood through a parallel with the Fermi problem [15-17] which is concerned by the dynamics of a single ball in a vibrated 1-d cavity with no loss: Indeed the Fermi case considers a no-loss condition and a periodic excitation; it predicts a random motion with no intrinsic period, with a mean transfer of momentum equal to 0, otherwise the energy of the ball would become infinite. On the other hand, the Fermi case has a periodic solution obviously, that consists of periodic impacts at phase $\varphi=0$, ($z_p=b$, $v_p=0$) with an adequate ball speed $v_{ball}$, corresponding to $<L>/L_{max}=1$; but this periodic orbit is not observed, hence it is likely unstable. So consider now a real case for which losses are non zero; consider now the limit of tiny losses and periodic orbits; this periodic orbit approaches the ideal Fermi case when phase $\varphi \to 0$ and $<L>/L_{max}=1$, with the speed $v_{ball} \approx 4\pi(L-d+2b)/\omega$; but it is shown above that this orbit is unstable; so this modelling predicts that periodic limit with $\varphi \to 0$ is unstable. This is just what is observed in Fig. 4.

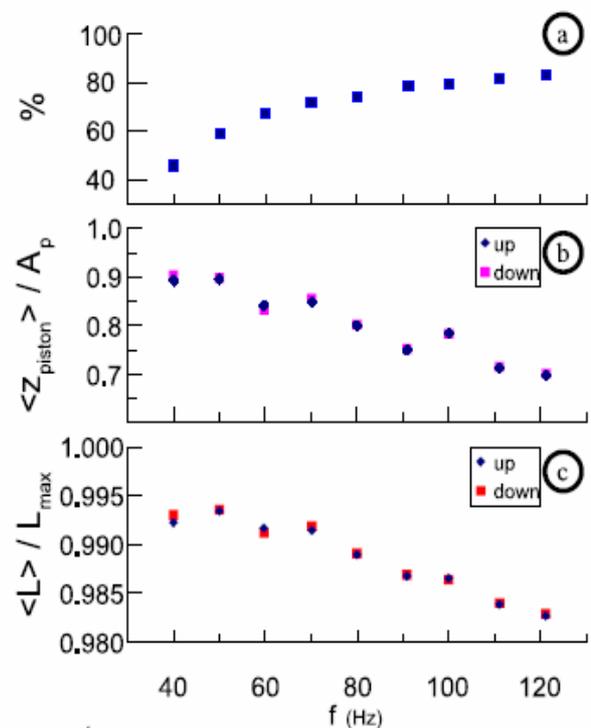

**Figure 4:** (a) Resonant rate (%); (b) Relative piston position; and (c) Relative cavity length, as a function of frequency, where $L_{max} = L-d + A_p$. $A_p=b$ is the amplitude of vibration. In part b and c, the up and down trajectories of the ball are hardly distinguished at this scale. (from [9])

♦ ***Comparison with results from 1-d simulations:*** By contrast, intermittency of the synchronous motion has been observed in 1-d simulations using an event-driven algorithm [20], see also Appendix C below. In some cases, one observes rapid variations of the relative standard deviation of the speed with the ratio b/L in simulations. This ratio depends on the initial condition. Hence, the stability of the periodic motion has to be studied in conjunction with the period of motion compared to the period of excitation; and its sensitivity to fluctuations has to be found as a function of the initial conditions in more details.

♦ ***Experimentally testing the stability:*** In the same way, some experimental test of the stability of the regular experimental trajectory of the ball has been performed forcing some change of boundary: for instance, in order to tempt destabilizing the impact position or the trajectory, i) different shapes of the surface of the piston basin have been used, ii) the cap surface of the sensor has been tilted up to 10° from the perpendicular direction of the vibration, iii) a second ball of lower diameter has been added; iv) up to 5 balls of 1 mm diameter have been introduced in the cell at the same times. Nevertheless for all these situations, the cyclic orbit parallel to the vibration direction has remained often stable, which confirms the loss of the ergodicity of this 3-d experimental system.

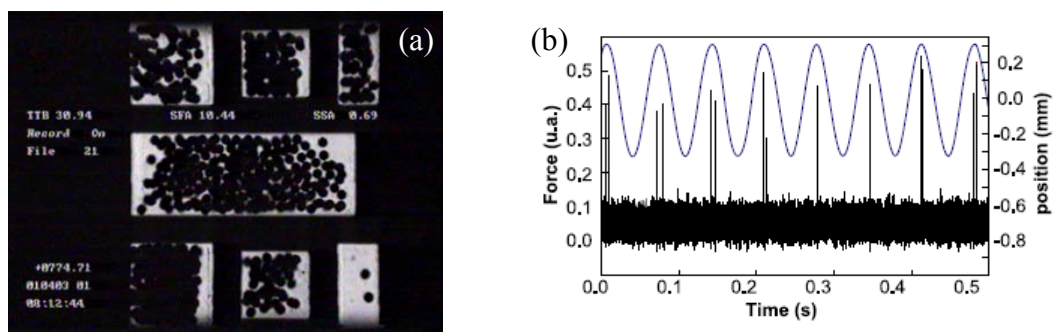

**Figure 5:** Maxus 5 experiment (10.5Hz, 3mm): The bottom right cell of (a) contains 2 balls (d=1mm) which move in phase and whose impacts with a sensor has been recorded (b)

♦ ***Maxus 5 experiment and the dynamics of a 2-balls system:*** In the micro-g experiment of Maxus 5 (launched from Esrange on April 1[st], 2003), these experiments have been repeated in one of the cells (the bottom righ cell of Fig. 5); the frequency and the box size were smaller (6Hz<f<60Hz, L=5mm, cross section S=10mm*10mm) and the whole box was moving. The cell was containing 2 spheres of diameter d'=1mm. In the micro-g experiment of Maxus 5, intermittency has never been observed, during the whole flight; the frequency and the box size were smaller (6Hz<f<60Hz, L=5mm) and the whole box was moving. This proves better resonance conditions, (or more stable conditions of resonance); this is perhaps due to improved technical characteristics, but it corresponds also to a relative increase of losses. Indeed, this experiment benefits from a better micro-g environment and/or it uses a different vibrator, working with a crank and a connecting rod. Still in this case, one observes some discrepancy with a periodic perfectly synchronous motion: as shown in Fig. 5.a, the bottom right cell of the Maxus 5 experiment contains 2 balls which moves approximately in phase and whose impacts with a sensor have been recorded, *cf.* Fig. 4.b. Indeed, these impacts are not strictly concomitant, that proves non perfect periodicity. I have no complete understanding of this effect at the moment; is it linked to some non local sphericity of the ball, to some dependence of the restitution coefficient to local variation, or to some local imperfection of the box? In the two first case at least, this would tell that 3-d dynamics should be always required to capture the effect of fluctuations.

♦ ***Parallel with classic billiard:*** In classic billiard theory applied to statistical mechanics, the real shape of the cavity plays a significant role on the ergodicity/non-ergodicity of the problem. In particular, billiards having the shape of a sphere cap should not present stable orbits. It would be interesting to demonstrate that cavity with peculiar shape would result also in the destabilisation of the resonant trajectories, hence improving the ergodicity of the dynamics. In the same way, using non spherical particles, adding some important surface roughness or adding some hard convex fixed obstacles into the cavity should also improve the "quality of the ergodicity". This is why perhaps the existence of a few balls, instead of a single one, changes the qualitative nature of the dynamics, as soon as the probability of ball-ball collision becomes large enough. This forces the problem of granular gas to be rather ergodic. However, this remains to be demonstrated.

♦ ***Subsonic nature of the excitation:*** At last, an other remarkable experimental result of the 1-ball experiment is that the speed $v_z$ of the ball is always larger than the piston speed $b\omega$. In the present terminology, it means a "subsonic" kind of excitation. The ratio $v_z/(b\omega)$ is much larger in the case of resonance than within erratic motion. This is in agreement with 1-d simulations and theoretical description [20]. This shows also that the "supersonic" excitation observed when a collection of beads fills the cell is linked to the dissipation during ball-ball collision. As already recounted, granular gas exists only in quite dilute conditions, *i.e.* the so-called Knudsen regime, in the limit of fixed L and $\rho \rightarrow 0$, so the case of few particles in interaction will be investigated in order to investigate the "catastrophic" influence of dissipation by grain-grain collision in granular-gas mechanics. A conclusion of the present section is also that the restitution coefficient $\varepsilon = -v_{out}/v_{in}$ of a frontal collision is rather large and remains rather constant, *i.e.* $0.9 < \varepsilon = -v_{out}/v_{in} = 0.95$, all along the experimental range of $v_{in}$, *i.e.* $v_{in} < 2.5 m/s$; this is in disagreement with previous attempt of simulations, that required fast dependence of $\varepsilon$ on $v_{in}$, *cf.* Fig.2.

♦ ***Concluding remarks:*** the dynamics of very loose granular gas has been investigated when particles do not anymore interact with one another. It has been found (i) that the behaviour of such a gas confined in a closed box can be extremely non ergodic, (ii) that its dynamics is dominated by the coupling with the boundaries and (iii) that it is controlled by the effect of solid friction which couples translation and rotation during collisions. This study shows also that the behaviour of a dissipative billiard is quite different from a non dissipative one.

It is remarkable that the coupling between rotation and translation during collisions is often neglected in many simulations. This makes the domain of *ab initio* numerical technique little predictive at the moment and limited to interpolation in between experimentally tested situations. It is rather strange that no numerical studies have tend to define the limit of the domains of applicability of the collision laws they used, because it does not require so many grains and so much computer time. This experiment tells an urgent need of testing the codes and of changing the protocols.

The dynamics observed here cannot correspond to the case of the so-called Knudsen-like regime, for which particles explore ergodically the space. I conclude that the understanding of the low density limit of non-interacting granular gas should be revisited, in the absence of gravity, in order to investigate the regime of "small-excitation" and $V_{piston}/v_{ball} \approx \pi b_p/(L-d)$, and the cross-over to a non periodic regime for which the trajectory should remain linear and 1-d, and $v_{ball}$ larger than $b\omega$. It can permit also to check the relevance of a threshold value, *i.e.* $\pi b/(L-d)= (1-\varepsilon)/(1+\varepsilon)$, above which the dynamics of the single ball is periodic in a vibrated box.

## 5. The case of a granular gas in Knudsen regime

When increasing the number of particles in the box, collisions between particles occur. This randomises the motion and slows down the dynamics. The motion looks much more erratic than the one with a single ball. However, does it really behave as a gas? Can one define a temperature of the system [7]? How this temperature scales with the motion parameters of the box, *i.e.* b and cos(ωt), with L? Does the "temperature" vary with the number of balls in the cell?

Part of the answer has been given by the Mini-Texus 5 rocket flight, from which few features have been demonstrated; these are:

(i) the supersonic nature of the excitation of the container, *i.e.* the typical speed bω of the container is much faster than the typical speed of the grains in the "gas" state,

(ii) the gas state seems to exist in the Knudsen regime, since as soon as the mean free path $l_c$ becomes smaller than the container size L, a cluster forms.

(iii) the distribution of impact strength seems to obey an anomalous scaling, even in the gas state.

So a series of experiments have been performed with the Airbus A-300 0-g of CNES to investigate the statistical mechanics property of the gas state at low density. Indeed, the level of micro-gravity condition, *i.e.* Δg=0±0.01g, is too poor in the aircraft to allow the correct investigation of the cluster dynamics because the cluster may explode rapidly on the wall after being formed, due to the erratic motion of the container: as the typical cell size is L=0.01m, the typical coherence time $\tau_{cluster}$ of the cluster can be estimated to $\tau_{cluster}=(2L/\Delta g)^{1/2} =0.4s$.

## 5.1. Experimental results:

Fig. 6.a reports a typical recording of the signal from the force sensor; it exhibits a series of peaks of different amplitude, labelled I here after, at different instants. Each peak looks as the blue curve in Fig. 1 and it corresponds to an impact. The experiment consists to study the number of collisions as functions of the number of balls in the cell, of the piston speed $V=b\omega$, of its frequency $f=\omega/(2\pi)$ and of the amplitude of vibration. Each parabola gives 20 s of milli-gravity condition, but the statistics are issued from the 16 central seconds, in order to limit noise.

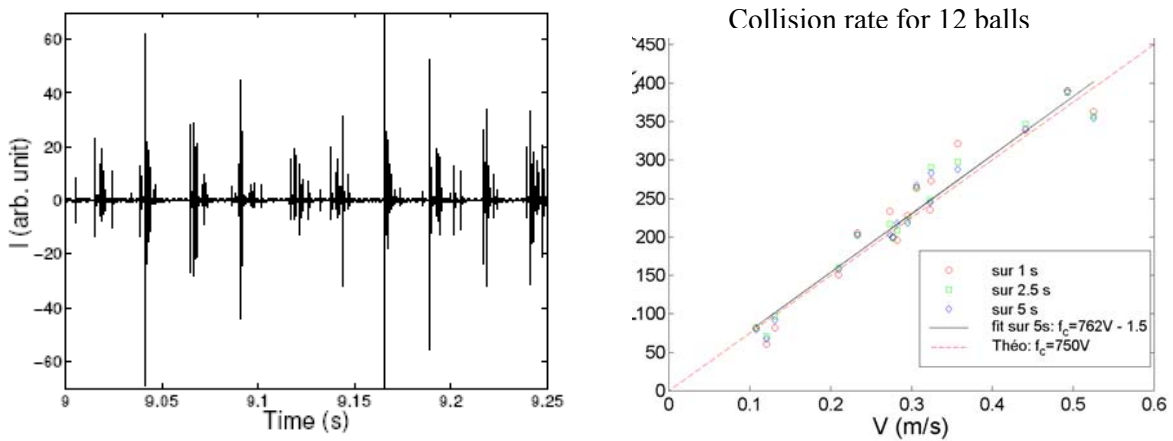

**Figure 6: 6.a:** Typical time dependence of the signal (of maximum amplitude I) from the force sensor recorded during 10 periods(=0.25s) of vibration showing 106 collisions. The parameters of vibration are $f=\omega/(2\pi)= 40$ Hz, A = 1.96 mm. The number of balls in the cell is : N = 12; $d_{ball}=2\pm0.002$mm, $D_{cell}=13.0$mm, $H_{cell}=10$mm

**6.b.** Typical variation of the number of collisions per second, averaged over different periods of times (1s, 2.5s 5s), for 12 balls, as a function of the piston speed $V=b\omega$ .
From [11]

Fig. 6. b gives the collision rate, *i.e.* the number $N_c$ of total collisions with the sensor divided by the duration of the measurement, in the case of a cell containing 12 balls at different frequency and amplitude. It shows that the controlling parameter is the piston speed $V=b\omega$. The behaviour is linear. It means that the typical speed of a ball scales linearly with the piston speed. As Fig. 7 demonstrates it, this trend is independent of the number of balls in the cell, within the limited range of b and $\omega$, and for the different number N of balls contained in the cells, *i.e.* 12<N<48. So the average speed of the balls $<v_{ball}>$ scales linearly with $b\omega$. Some of the values of the experimental parameters investigated are listed in Table 1.

One observes also from Fig. 7 that the number of collisions $N_c$ does not scale linearly with the number N of balls in the cell but follows a scaling of the kind $N_c \approx N^{0.6}$. If the dynamics of the particles were independent, $N_c$ should scale as N. The law obtained here shows on the contrary that the mean speed $<v_{ball}>$ of the particles scales as $b\omega/N^{0.4}$, *i.e.* $<v_{ball}> \propto b\omega/N^{0.4}$.

The roundtrip of a ball lasts approximately $2(L-d)/<v_{ball}>$; hence the numbers of collision is $N_c \approx N$ T $<v_{ball}>/(2L)$. This leads to $<v_{ball}>=2LN_c/(NT)$. And to the data of table 1. The typical relative standard deviation $\Delta k/\langle k \rangle$ computed on $k=\langle v_{ball} \rangle/(b\omega)$ is about 0.1, spanning from 0.07 in the case of N=36 to 0.16 in the case N=48.

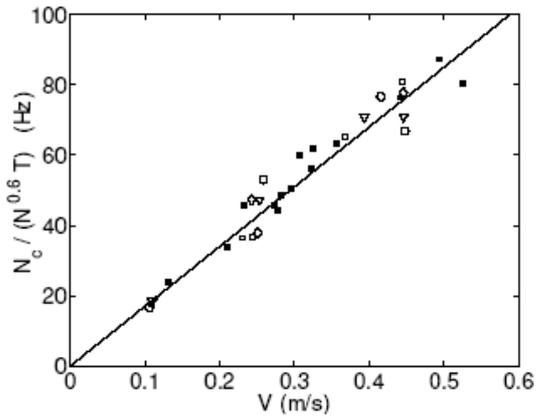

**Figure 7:** Total number of collisions $N_c$ observed during time T, rescaled by $N^{0.6}$ and by time, $N_c/(T\ N^{0.6})$ as a function of V=b$\omega$ for N =12 ($\square$) and ($\blacksquare$); 24($\lozenge$); 36($\triangledown$); 48($\circ$) during T = 16 s of low gravity.
$\blacksquare$-marks are from a previous set of experiments at fixed N = 12 and 15 different velocities which are not listed in Table I.
N is the number of particles in the cell.
Solid line corresponds to the fit $N_c/(TN^{0.6}) = \alpha V$.

| # | N | b (mm) | f=ω/(2π) (Hz) | V=bω (m/s) | Γ=b²ω²/g | $N_c$ | k = $\langle v_{ball} \rangle$/bω |
|---|---|---|---|---|---|---|---|
| 1 | 12 | 0.92 | 40 | 0.23 | 5.9 | 2591 | 1.17 |
| 2 | 12 | 0.65 | 59.7 | 0.24 | 9.3 | 2605 | 1.13 |
| 3 | 12 | 0.88 | 80 | 0.44 | 22.7 | 5735 | 1.36 |
| 4 | 12 | 0.64 | 90.9 | 0.37 | 21.4 | 4617 | 1.30 |
| 5 | 24 | 0.96 | 40 | 0.24 | 6.2 | 5097 | 1.11 |
| 6 | 24 | 0.67 | 59.7 | 0.25 | 9.6 | 4078 | 0.85 |
| 7 | 24 | 0.88 | 80 | 0.44 | 22.8 | 8362 | 0.99 |
| 8 | 36 | 0.44 | 40 | 0.11 | 2.8 | 2538 | 0.80 |
| 9 | 36 | 0.67 | 59.7 | 0.25 | 9.7 | 6496 | 0.90 |
| 10 | 36 | 0.88 | 80 | 0.44 | 22.8 | 9744 | 0.77 |
| 11 | 36 | 0.69 | 90.9 | 0.39 | 22.9 | 9741 | 0.87 |
| 12 | 48 | 0.42 | 40 | 0.11 | 2.7 | 2728 | 0.65 |
| 13 | 48 | 0.69 | 59.7 | 0.26 | 9.9 | 8650 | 0.87 |
| 14 | 48 | 0.89 | 80 | 0.45 | 22.9 | 10906 | 0.63 |
| 15 | 48 | 0.73 | 90.9 | 0.41 | 24.2 | 12512 | 0.79 |

**Table 1:** List of vibration parameters studied during A300 parabolic flights. Each parabola lasts 20 s of low gravity. N is the number of grains in the cell. V = 2πbf and Γ = 4π²b²f² /g are respectively the maximal piston velocity and the dimensionless acceleration of vibration. The number of collision $N_c$ with the sensor is detected during T = 16 s of low gravity to avoid transient states.

## 5.2. *Statistical analysis of experimental data:*

From the recording of Fig. 6, and from other analogue data, one can get access to other parameters of the statistics and check the validity of a series of hypotheses.

### • *Distribution of waiting times:*

For instance one can draw the statistics P($\Delta t$) of the waiting times $\Delta t$ in between successive impacts with the sensor. Indeed, as the gauge is fixed, this statistics may test whether the particles remember the excitation by the piston, in which case one expects to see some modulation at the frequency of impact; or if the particles are correlated in motion, in which case one shall observe a peaked distribution.

In the case of uncorrelated motion between particles, one shall expects a normal exponential distribution: be p($\delta t$) the probability that any particle hits the gauge during δt; within the previous assumption of randomness p(δt)=p$_o$δt, the probability that no hits occurs during Δt is then

$$P(\Delta t) \propto p^2[1-p]^{\Delta t/\delta t} \propto p^2\, e^{-p_o\Delta t} \qquad (5)$$

The normalisation of Eq. (5) shall be obtained by imposing the mean number of collisions per unit time p$_o$ as the inverse of the mean waiting time, *i.e.* <Δt>=1/p$_o$. Fig. 8 reports the distribution of waiting times in between successive impacts for the different number of balls studied. As they are exponentially distributed, the result is compatible with random uncorrelated independent motions of the particles. As shown by Figs. 6 & 7, the frequency, *i.e.* p$_o$, of collision varies linearly with the piston speed V=bω. Hence, p$_o$ shall be proportional to bω, which is found on Fig. 8.

♦ *Is p$_o$ related to some mean free path l$_e$?* One can use p$_o$ and bω to define a "pseudo mean free path" l$_e$=bω/p$_o$. This has been done in [11]. It seems now to me that it is not a good idea indeed, because p$_o$ scales linearly with the gauge surface; hence it might vary just by varying the sensor size.

In other words, as the dynamics involves collisions of particles with one another and with the box, its physics is governed by the ratio L/l$_c$, that remains constant when the number n$_l$=Nd²/D² of layers is kept constant; this means also a constant mean ball speed. Changing d at constant n$_l$ does not change the dynamics but changes the collision frequency with the wall p$_o$; hence l$_e$ is expected to vary according to:

$$l_{e1}=l_{eo}d_1^2/d_0^2$$

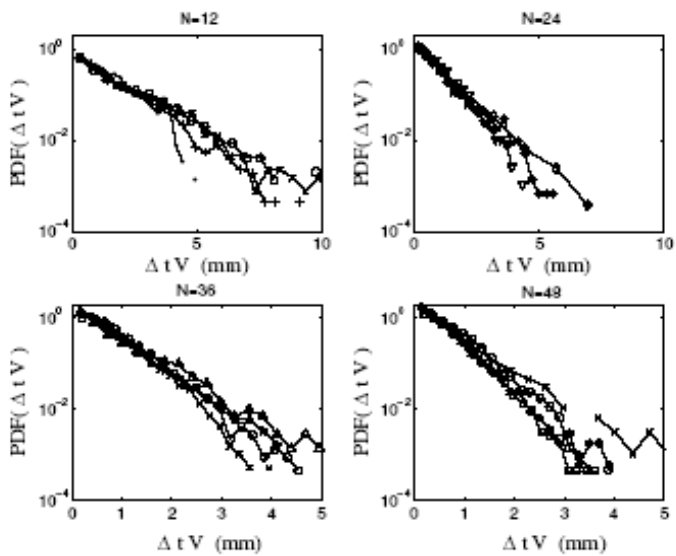

So p$_o$ is just related to a mean waiting time; it scales linearly with the gauge surface, with the number of particles (at constant n$_1$) and with their speed; within this basis of parameters, it does not depend on the ball diameter.

♦ *Question : Is the flow against the gauge modulated and periodic?* Indeed, one sees on Fig. 6 some modulation of the flow at the frequency equal to the excitation frequency about. On the other hand, curves of Fig. 8 do not exhibit any modulation, while the timescale of Δt spans over a period of vibration, *i.e.* 2π/ω, in many cases. This is the case for (N=12 curves 1-4), (N=24,curve 7), (N=36, curve 8), (N=48, curve 12). Furthermore the other durations range around 2π/ω. So a question remains: Why is a modulation not observable in Fig. 8? This seems to cast some doubt about the validity of the data.

- *Distribution of impact amplitude* **I** *during collisions with the gauge:*

One can study also the distribution of impact intensities I. This is done in Fig. 9 for the 4 different numbers of balls in the cell, *i.e.* N=12, N=24, N=36 and N=48. Here I is the maximum amplitude of the signal of Figs. 1, 5 or 6, at each impact. *A priori*, as the gauge is fix, one expects that I is proportional to the transfer of momentum $(1+\varepsilon)mv_{ball}$, with $\varepsilon$ being the restitution coefficient. As $v_{ball}$ is proportional to V=bω the horizontal scale of I has been divided by bω to get rescaling. This is indeed satisfied, since the different data fall into a single trend for each N; hence, the rescaling used is correct.

Consider now the experimental data of Fig. 9 which are issued from the three sets of curves at N=24, N=36 and N=48. They show that the probability distribution p(I) follows an exponential decrease as a function of I/(bω), with a rate that depends on N. On the other hand, the data from the case N=12 shows up some deviation from this single-exponential trend.

In practice one expects I to be proportional to $mv_{ball}$. Be $f(v_{ball})dv$ the probability of finding a ball with speed v in the vibration direction in a volume element $d\varpi=Svdt$, the number of impacts within $dv=Svdt$ shall varies as:

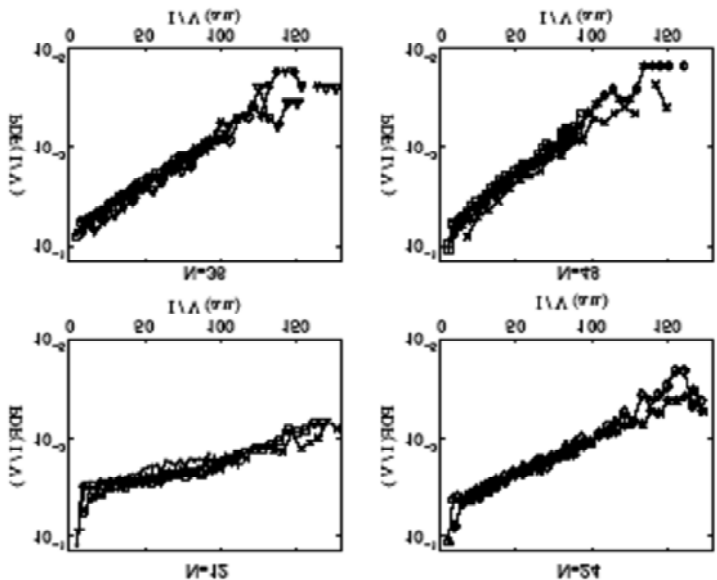

$$p(I)dI \propto dN_{impact} = v_{ball}\ f(v_{ball})\ dv_{ball}\ Sdt \qquad (6)$$

So, if I is proportional to $v_{ball}$, Eq. (6) predicts that f(v) scales as p(I)/I, or in other words as p(v)/v:

$$f(v_{ball}) \propto (1/v_{ball}) \exp\{-v_{ball}/v_o\} \tag{7}$$

♦ **Problem with Fig.9 data:** So Eq. (6) predicts that the number of particles having a very small speed diverges as $1/v$. In other words it predicts that most of the particles remain at very low speed while one or two of them are quite speedy. This is quite interesting since it predicts anomalous correlation in the motion. However I believe it carries some other problems, that makes this solution untenable:

Indeed, in the experimental case, *i.e.* N ≥12, one observes that beads have erratic motions, that implies collisions between the balls and random transfer of momentum. One can evaluate the mean free path $l_c$ of a ball; it is $l_c=(D/d)^2L/(\alpha N)$ where $\alpha$ is a coefficient ranging from 1 to 4 about (see the discussion below). It means that $l_c$ is of order 3L at most. The above solution requires that some of the beads have large speeds, which means $v_{ball} \gg b\omega$. Hence it requires that some of the balls hits the piston efficiently during a series of impacts without transferring their energy to the other beads in between, hence without having hit those beads; as the momentum transferred per hit with the piston depends on the ratio $b\omega/v_{ball}$, but cannot overpass $2mb\omega$, and as the length of a round trip is 2L, the speed $v_{ball}$ scales approximately as $2nb\omega$; the probability that such an event exists is linked to the survival probability that a ball has not collided with any other ball on a travel of length 2nL; hence it decreases as $\exp(-2nL/l_c)= \exp[-vL/(l_cb\omega)]$. This leads to the predicted trend:

$$f(v) \propto \exp[-vL/(l_cb\omega)]=\exp[-v/v_o] \tag{8}$$

with $v_o=\alpha(d/D)^2Nb\omega/L$; this seems the correct experimental scaling at large I. So it predicts that the slope increases by a 2 factor when the bead number passes from N=24 to N=48. Data of Fig.9 gives the ratio 1.7 about, instead of 2. Also the range predicted for $v_o$ is $b\omega/4<v_o<3b\omega$ , which is correct.

So, this may be an explanation of the exponential tail at large speed. However the problem remains to understand why most of the particles have a very little speed, why a numerous part of them are not excited by collisions with high-speed particles and why the pressure is carried by a very small amount of beads? In other words, if $f(v)$ is given by Eq. (7), its integral diverges as $\log(v)$ when $v \rightarrow 0$. This is not possible because of the finite number of particles. So it is tempting to try a second explanation.

## 5.3. Introduction of the transfer function of the transducer  r(I,v):

In the preceding section, *i.e.* §-4, the dynamics of a single ball has been studied; and the motion of the ball has been found 1-dimensional and quasi periodic; it is synchronised on the piston movement, with a speed $v_{ball}$ controlled by the frequency $f=\omega/(2\pi)$ of the excitation and by the size of the box L, according to $v_{ball} \approx \omega(L-d)/\pi$. This experiment can be (and has been) used to calibrate the gauge response I to the impacts. This has been done and it has been found that the relationship between I and $\omega$ is rather linear; this requires however to take  some caution is taken (it is better to use a thin "rigid" lid in front of the gauge for instance and to force the impacts to be localised at a well given position). During the calibration experiments, the response has been obtained mainly in the centre of the gauge surface; this was easy to obtained with some geometry of the lid, because the motion of the ball is very regular in the case of the 1-ball system under periodic excitation.

On the contrary, the system investigated in the present experiment exhibits quite erratic 3-d behaviour; and the position of the successive impacts may be distributed at random on the whole surface of the gauge. Furthermore, the sensor was not mounted with a thin lid on top of it. So, if the sensor response depends also on the precise position of the impact on the gauge surface, then one shall expect a response with a given distribution. Such an effect can be observed on Fig. 5, where a series of impacts are reported for a system of 2 balls travelling synchronously with approximately the same speed; but the amplitudes of I varies erratically from an impact to an other one.

So, be r(I,v) the statistical distribution of the response I of the sensor to the impact of a ball at a given speed v in the direction perpendicular to the gauge surface (and parallel to the vibration excitation). Be f(v) the distribution of speed of the balls. The number of collisions with the gauge is proportional to v f(v). Hence one expects the sensor response to be:

$$S(I) = \int v\, f(v)\, r(I,v)\, dv \qquad\qquad (9.a)$$

Suppose now that r(I,v) is broadly distributed, limited by I=0 and by I=mv. A typical example of such a brod distribution canl be the gate function G(I,0,mv) starting at I=0 and ending at I=mv. *i.e.*

$$r(I,v) = G(I,0,mv) = \Theta(I,0)[1 - \Theta(I,mv)] \qquad\qquad (9.b)$$

where $\Theta(y)$ is the Heaviside function starting at x=y (*i.e.* $\Theta(x.y)=0$ for x<y and =1 for x>y). As r(I,v) is a probability distribution its amplitude shall scale as 1/(v). So applying this distribution in Eq. (9) leads to:

$$S_G(I) = \int_{I/m}^{0} f(v)\, dv \qquad\qquad (10)$$

In this case one can get the distribution r(v) directly from experimental data, just by derivation:

$$f(v) = - dS_{exp}(I)/dI \qquad\qquad (11)$$

♦ ***Why the response function of the gauge is large:*** The way the sensor is manufactured can explain why its response r(I,v) is so broad:  it is a thin elastic membrane which is deformed by the impacts and which is attached at its boundary; the deformation of the membrane is measured.  So the response to an impact located on the centre shall be much larger than the one located near a boundary.

### 5.4.  *Fig. 9 results revisited:*

Interpreting the results of Fig. 9 within this scheme leads to the following features:

♣ The top left figure of Fig. 9 shows an $S_{exp}(I)$ that does not follow a strict exponential: approximately, it starts as a rather flat curve followed by an exponential decrease. In the framework of a response function of the kind of Eq. (9.b),  this implies a very little probability of getting particles with low speed, till a given value $v_1=I_1/m$, *i.e.* $I_1 \approx 50\text{-}100$ in the a.u. used in Fig. 9, then f(v) decreases exponentially. So in a first approximation one gets the following trend:

N=12 :   f(v)=0     for v<$v_1$=50−100 a.u.                                      (12.a)

N=12 :   f(v)=exp(-v/$v_o$)  for v>$v_1$=50−100 a.u.                              (12.b)

with $v_o = \gamma_{N=12} b\omega$

In practice, the discontinuity is likely less efficient and f(v) has probably the following shape: it starts at f(v=0)=0, then it grows between v=0 and v=$v_1$; then it decreases exponentially. From Fig. 9, one sees also that the shape of f(v) in between v=0 and v=$v_1$ depends on the parameter of vibration, while it does not depend at large speed. This may imply some effect of coupling between the dynamics of few particles at small speed.

♣ On the other hand, in the three other figures, corresponding to N=24, 36, and 48 balls, $S_{exp}(I)$ decreases exponentially from v=0; hence f(v) follows the distribution :

N=24, 36 or 48:   f(v)=exp(-v/$v_o$)    with $v_o = \gamma_N b\omega$                        (12.c)

$\gamma_N$ is a parameter which depends on N, but not on b and $\omega$, as experiments tell.

♦ *Explanation of the exponential tail of* **f(v)**: A possible explanation for the exponential decay has been already given in a previous sub-section, *i.e.* §-5.2. It is based on the fact that a ball can gain only some finite discrete amount of speed after each collision with the piston,  typically $\Delta v < 2b\omega$; but it risks also to loose its own kinetic energy if it meets an other ball, *i.e.* the loss of speed is proportional to v in this case. Hence be n the number of round trips a ball performs in between two collisions with an other ball, its speed scales approximately as v=2nb$\omega$ and the probability of such an event varies as exp(-2nL/$l_c$); replacing n by v/(2b$\omega$) leads to the approximate exponential trend, *i.e.* exp[-vL/(2$l_c$b$\omega$)], where $l_c$ is the mean free path. Hence this explanation fits the experimental data.

Two problems remain however: (i) what is the precise value of $l_c$? (ii) does really the gain vary linearly with n? Does it depend or not on the ball speed?

• *Problems of mean free path $l_c$ and of collision cross-section:*

Previously, it was assumed that the mean free path varies as $l_c$=(D/d)²L/($\alpha$N), where $\alpha$ is a coefficient ranging from 1 to 4 about. Indeed, $l_c$ is related to the collision cross-section, which is assumed to be $4\pi d^2$ in the classic theory of gas, which leads to $\alpha$=4.

However this value may be slightly different here because the real cross-section depends on the precise problem; here the problem is more the loss of energy than the redistribution of energy:

So, consider first very eccentric collisions; in this case the speed component which is normal to the contact is smaller than the tangent one, so that the solid friction force is small and the dissipation due to solid friction is small; in the same way, as the restitution coefficient is large, the dissipation is small; this reduces in turn the effective cross-section. But, eccentric collisions redistribute also the speed in the other different directions. On the other hand, because of the large restitution coefficient, central collisions dissipate mainly the kinetic energy which was stored in the rotation degrees of freedom; these collisions redistribute also partly the speed direction. This leads to define an effective cross section area in between $\pi d^2$ and $4\pi d^2$, in the present case; but it is a crude approximation, which depends likely on different parameters such as the number of balls, the real energy stored in the system in the rotation degrees of freedom …

•  *Problem about the speed gain* $\Delta V=(v_{out}-v_{in})/v_{in}$:

In the same way, the gain of speed after each collision with the piston is not single valued; it depends on the phase of the period at which the collision occurs and of the speed of the ball. Even on average, it depends then on different parameters:

The first parameter is the spatio-temporal structure of the flow. Is it modulated in time or not? Indeed since the excitation is modulated, this may generate a flow which is modulated in time and space and which propagates along the cell; a typical example has been observed in section 4, where a synchronous motion has been observed. In order to detect such an effect, one should check the modulation of the frequency of the collisions with a fix gauge: this rate should exhibits some modulation within coherent phase with the excitation; it would lead also to a modulation of the time in between two successive collisions.

Nevertheless, when no modulation exists, one can assume a continuous flow; but as the piston moves periodically, its probability of collisions with the balls is modulated. The problem can be solved using the random phase approximation [20] in the limit of large cell; it works as follows.

♦  *The random phase approximation (RPA)*: the flow is assumed to be continuous with some distribution of speed f(v) in the lab frame; and the time needed to the ball to travel across the cell is considered to be long compared to the period of motion of the piston. Since the piston speed depends on time the probability of collision and the transfer of momentum depend both on time [20].

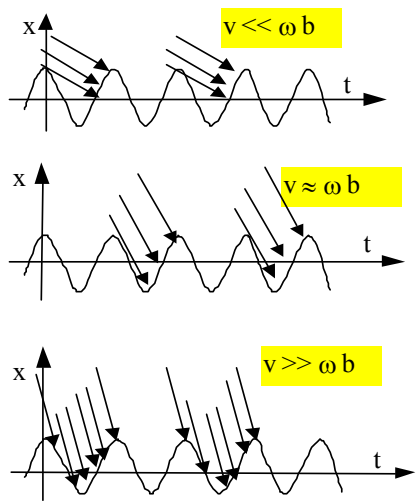

*Figure 10: Comparison of ball and wall motions for different values of v and bω when v<bω, some screening is obtained that forbids ball-wall collisions during some part of the period. When v>bω, no screening is obtained, but the probability of ball-wall collision still depends on time since it is proportional to v-bω sin(ωt).*

Be $t_{coll}$ the time at which collision occurs, the probability of collision is given by:

$P=[v_{ball}-b\omega \sin(\omega t_{coll})] f(v)$          with $v_{ball}-b\omega \sin(\omega t_{coll}) >0$

$P=0$                 if $v_{ball}-b\omega \sin(\omega t_{coll})<0$

Hence it depends on the collision time and of $v_{ball}$. In the same way, be $\varepsilon$ the restitution coefficient the gain in speed $\Delta v_{ball}$ is given by:

$$\Delta v_{coll} = (1+\varepsilon) \, b\omega \, \sin(\omega t_{coll}) +(1-\varepsilon) \, v_{ball} \qquad\qquad (13)$$

whose upper limit is $2b\omega$ , and is obtained for $\varepsilon=1$; but the gain varies with $t_{coll}$.

In the RPA, one is looking for a non periodic solution, with large enough fluctuations of speed so that there is a large indeterminacy for the next collision time, *i.e.* $\delta t>T=1/f$; as one is looking also for stationary solutions, one shall impose a gain $<\Delta v_{coll}> = 0$ in average . This last case has just been studied in [20] under a mean field condition. What occurs is the following:

♣ As exemplified in Fig. 10, when the considered speed $v_{ball}$ becomes smaller than $b\omega$, collisions cannot occur during some part of the period; and this part of the period increases when the considered speed $v_{ball}$ becomes smaller and smaller. (This forbids almost completely to get collisions when the piston moves backward).

♣ When $v_{ball}>b\omega$, collisions occur when the piston moves backward and forwards, with different probabilities; it gives a net gain in average. When $v_{ball}>>b\omega$, the net gain becomes quite small to tend to 0 when $v_{ball}\rightarrow\infty$. This shows that the gain depends on $b\omega/v_{ball}$, $\Delta v_{coll} =-(1+\varepsilon) \, b\omega \, \sin(\omega t_{coll}) +(1-\varepsilon) \, v$.

♦ *periodic flow:* The other way to solve the problem is to look for rapid periodic motion, with steady conditions; in this case, one expects $\Delta v_{coll} =0$; so, in the case of a moving cell, this implies $v_{ball}= (1+\varepsilon)b\omega \sin(\omega t_{coll})/(1-\varepsilon)$. And the complete periodic solution is found by imposing that the time of flight for a half roundtrip is equal to half the period of excitation: $[L-d+2b\cos(\omega t_{coll})]/v_{ball} =\pi/\omega$. It is the case studied in §4. But it does not concern this section for which the flow is not periodic.

### 5.5. A simple model « à la Boltzmann » for the f(v) distribution:

I derive now a simple model which allows to get the natural trend of Fig. 9, for $N\geq24$. It comes from the remark that the action of the piston when it moves is to tends to impose some speed in a given direction, rather than a kinetic energy; the role of the collision with other particles is to tend to redistribute the speed into different directions due to collisions.

One notes also that according to the collision rules, the total momentum is preserved after the collision, but not the total energy. Hence, coming back to the Boltzmann's equation of section §3 and its resolution, it is tempting to consider the system of gas to be coupled to a velostat, which preserves momentum, rather than a thermostat. Hence, it is tempting to consider that the quantity whose disorder has to be maximised is the speed in the direction of vibration.

This allows to proceed as in classic statistical physics [23]: one assumes that the observed state is the most probable one; this means that it is the state with the largest possible complexion number W; so W shall be optimum, and ln(W) too. However, it shall also correspond to a possible state of the system. As the system is coupled to a velostat, its mean speed v is well defined. Be $n_i$, the number of particles with speed $v_i$, the number of complexion W is $W=N!/(n_1!,…n_i!..)$. The other above conditions give:

$$N=\Sigma_i n_i \tag{14.a}$$

$$Nv = \Sigma_i v_i n_i \tag{14.b}$$

Or as dv=0, dW=0, dN=0, and using the Stirling relation $lnN! \propto NlnN-N$, these 3 conditions can be rewritten:

$$d(lnW)=0= -\Sigma_i ln(n_i) dn_i \tag{15.a}$$

$$0= -\Sigma_i dn_i \tag{15.b}$$

$$0= -\Sigma_i v_i dNi \tag{15.c}$$

A way to condense these n=3 conditions into a single one is to use the technique of the Lagrange multipliers and to introduce n-1 coefficients $\lambda_1$, $\lambda_2$ , …, called the Lagrange multipliers

$$0= -\Sigma_i \{ln(n_i) - \lambda_1 - \lambda_2 v_i \} dn_i \tag{16}$$

where $\lambda_1$ and $\lambda_2$ are now fixed by the experimental conditions and the $n_i$ are uncoupled to one another; this leads to the general solution:

$$f(v)= A \exp[-\lambda_2 v] \tag{17}$$

which is just the right behaviour found in Fig. 9, with $v_o=1/\lambda_2$ !

Hence everything occurs as if one could treat the system of grains as coupled to a velostat. This confirm recent assumption [7]. The problem which remains is that the typical speed of the ball in the cloud depends on the mean free path compared to the cell size, or in other words of the number of particles N. Hence it means that the velostat is not a true velostat, since it imposes a typical speed which depends on the number of balls. This leads to the following remarks:

♦ *Remark 1:* The above model is *ad hoc*, since the real correct reasoning to get the distribution function f shall proceeds as it is explained in section 3. Hence, it still requires to transform Eq. 2 into Eq. 3; this requires in turn the preservation of both momentums and energy; this is not the case in granular systems.

♦ *Remark 2:* One shall note also that both distributions f(v)=exp(-v/v₀) and f(v)=ex{-½mv²/(kT)} are steady solutions of the time independent Boltzmann's equation when the collisions satisfy both momentum and energy preservation. So both solutions exists for a classic gas; but the making or manufacturing of a velostat working for classic gas is not usual, and probably not known.

♦ **Remark 3:** Anyhow, according to the above experimental results that concern few balls (and as far as these results are correct), the correct solution of the steady Boltzmann's equation with dissipation due to collisions should be an approximate exponential tail, *i.e.* f(v) ≈ exp{-v/v$_o$} , at least as a first approximation. It may then happen that the solution of the **dissipative gas** may be approached starting from the one of a **non dissipative** gas in a velostat, that implies f(v)=exp{-v/v$_o$}, and can be computed by perturbation, with the perturbation being the dissipating collisions.

However, this is not obvious, and remains to be demonstrated, because the single particle problem does not lead to an exponential distribution of speed.

## 6. Discussion based on an attempt of solving the Boltzmann's equation in the case of dissipative systems:

The aim of this section is not to solve the system of equation leading to the steady distribution function obeying the Boltzmann's equation, but just to recall the different conditions which shall be written. Indeed, the distribution function evolves due to ball-ball collisions and due to collisions with boundary, which is studied separately here after because they are produced at different locations. In order to get some more simplification, it is also assumed that the system is homogeneous across a section perpendicular to the vibration direction. This is far from being demonstrated.

### 6.1. Bulk evolution:

One starts from Eqs. (1) and (2), and looks for a stationary solution of Eq. (1). This leads to

$$\mathbf{v}.\nabla_r \ f(p,q,t) = (\partial f/\partial t)_{coll} \tag{18}$$

$$(\partial f/\partial t)_{coll} = \int dp_1 \int d\Omega \sigma(\Omega) \ [|v'-v'_1|f(p')f(p'_1) - |v-v_1|f(p)f(p_1)] = 0 \tag{19}$$

with the preservation rules : m =m', m$_1$=m'$_1$, p+p$_1$=p'+p'$_1$. These are the only preservation rules since the collisions do not preserve energy. Hence, according to Eq. (18) one expects the speed distribution evolves all along the box. But the distribution shall obey some preserving rules; for instance the total flow in a cross section shall be equal to 0, which writes.

$$\int v \ f(p,q,t) \ ds \ dp = 0 \tag{19.a}$$

ds is the element of surface of the cross section considered. It is worth noting that local flow can be non zero in case of convection, and that

$$f(p,q,t) \neq f(-p,q,t) \tag{19.b}$$

because dissipation slows down the speed of balls.

## 6.2. Collisions with the wall:

As the system is dissipative its motion shall be sustained. This is done from a source term which occurs at the boundaries. In 3-d collisions, the rule shall include rotation, normal restitution coefficient, and solid friction. The collision rules shall be written as some condition:

$$f(p',q,t) = \int g(p,t)f(p,q,t) \, dp = 0 \qquad (20)$$

where g includes the effect of wall motion, b cos($\omega$t). As the ball motion is rather erratic and as its typical speed is small when collisions with other balls cannot be neglected, one can use the RPA approximation [20]. However as the wall motion is relatively fast compared to the ball speed, it may occur that collisions with the moving walls are forbidden during some part of the period of excitation (see Fig. 10). This occurs when $v_z < b\omega$. Following [20], one finds that the laps of time [$t_2,t_1$] during each period [$t_2$, $t_2 + 2\pi/\omega$] for which collision of a ball of speed $v_z$ with wall can occur depends on $v_z$ and can be computed:

$$t_1 = (1/\omega) \, Arcsin(-v_z/[b\omega]) \qquad (21.a)$$

$$v_z t_2 + b\cos(\omega t_2) = \cos\{Arcsin[-v_z/(b\omega)]\} + (v/\omega) \, Arcsin[-v_z/(b\omega)] \qquad (21.b)$$

with for instance the 1-d collision rules: $v'_z = -(1+\varepsilon)b\omega \, \sin(\omega t_{coll}) - \varepsilon \, v_z$, where $\varepsilon$ is the ball-wall restitution coefficient (this collision rules are valid only in 1-d) .

When $b\omega > v_z$, no such effect of screening exists; but the collision probability between the wall and the ball still depends on the relative speed $v_z - b\omega \, \sin(\omega t)$; so, even with a constant flux of ball , *i.e.* even when f(p) is independent of time, the term $v - b\omega \, \sin(\omega t)$ generates a net transfer of momentum from the wall to the gas in mean, as soon as v is finite.

So, this kind of excitation shall always generate a periodic distribution f(p,q,t) near the moving wall. How deep this distribution modulation may propagate into the cell is not simple: for instance such periodic behaviour is frequently observed far above the bottom in 1-g experiments indicating that the density evolves with time and height.

So, in micro-gravity condition, one may ask if f is still modulated with time or not. For instance near the wall, one expects that there are an incoming flux $j = \int_{v>0} vf(v)dv$ of balls and an outgoing flux $j' = \int_{v<0} vf(v)dv$ , with two different mean speeds $<v> = \int_{v>0} f(v)dv$ and $<v'> = \int_{v<0} f(v)dv$ and different width distribution $\delta v$ and $\delta v'$. If the flow is stationary one shall have also $j + j' = 0$.

In absence of collision, the periodic oscillation of $\int f(v')dv'$ will be smoothened after a typical length $L_1$ such as $L_1 \delta v'/v'^2 \approx T = 2\pi/\omega$ , or $L_1 = 2\pi \, v'^2/(\omega \, \delta v')$ . Hence, if the cell size $L \gg L_1$, the forcing shall not be seen deep in the cell in average, while it can still be observed on the collisions rule at the moving walls; hence, it does not mean however that no time modulation is observable on f, since one shall still see a modulation of f(v',t) when selecting a range of v with sufficient accuracy. When collisions happen this randomises more efficiently.

Anyhow, as soon as $L_1 \ll L$, the modulation is no more visible in average, and a random phase approximation can be used to calculate the outgoing distribution $f_+$ from the incoming one $f_-$, as in [20].

In order to see the difficulty, it is worth to investigate the 1-d version of this problem.

## 7. Conclusion:

The physics of granular gas has been importantly investigated during these last 10-15 years, using many different approaches; the ones stand from a theoretical approach; some others from 2d and 3d experiments on earth and in weightlessness conditions; others use numerical simulations and confrontation between numerical calculation and experimental data.

But complete answers remain little in the literature at the moment. For instance if one is interested by the problem of simulations, one can get from the literature different procedures to simulate the gas of particles and its dissipation, starting with different collisions rules; one uses (i) simply a normal restitution coefficient, or (ii) two coefficients (a normal and a tangent restitution coefficient), while others (iii) include particle rotation, solid friction, …. But none of these works tend to define clearly the minimum ingredients to get the true limit behavior of a dissipative granular gas as it is encountered in experiments.

This article has tried to define few results and ideas which might be used to guide the research; all focus on the problem of dissipation. They have been obtained using three different approaches: (i) the theoretical problem has been addressed using the Boltzmann's equation. (ii) Then the dissipation due to collisions with boundaries has been studied experimentally. (iii) The effect of ball-ball collisions has then be investigated.  The results can be summarized as follows:

♦ *Boltzmann's equation*

Here the problem is formulated within the formalism of the Boltzmann's equation; and the ingredients which are required to get a Maxwellian distribution as a solution of this equation are recalled. This allows to show that these requirements are not satisfied in the case of granular gases due to the dissipation during collisions. Furthermore, as collisions dissipate fast, as fast as they redistribute the speed direction, it is not obvious that a local temperature has the classic meaning, since it cannot be sustained during an even-small while without excitation.

♦ *1-ball experiment: the case of a granular gas of non interacting particles*

The case of non interacting particles is then investigated experimentally. Its attractor is found 1-d, so it exhibits a strong reduction from the real initial phase space dimension (11-d). It is due to the coupling between translation-and-rotation degrees of freedom during collisions with the moving walls. Hence it is proved here with this 1-ball experiment that accurate *ab initio* calculation shall contain ball rotation. (Of course, one could argue instead that 1-d modelling is enough!)

It is rather strange that no numerical studies have found this result, because it does not require so much computer time; so, it means likely that no study has tried to define the limit of the domains of applicability, starting with as less approximation as possible and testing successive approximations. This means an urgent need of testing the codes and of changing the protocols.

Gravity has to play an important role also, whose effect on granular gas behavior should gain to be clarified; indeed results in 1-g and in 0-g are often thought as equivalent, but the timescale and confinement are strongly affected by gravity. For instance the time scale of a bouncing varies as $t=v/g$ while it scales as $L/v$ in 0-g; and the particles are confined on the bottom in 1g while confinement requires two opposite walls in 0g. This asks the effect of boundary conditions: the presence of a second walls is crucial here; are lateral walls playing similar importance?….

This experiment shows also that the single ball system is not so well known and defined. For instance take the case of the 1-d bouncing ball system in 1-g without rotation. Theoretical study has shown that it should lead to quasi-periodic orbits. Taking the experiment, one finds complete chaos. Are rotation degrees of freedom involved in this change of behavior?

Indeed, this study leads to reformulate the problem of a dissipative billiard; is it different from a normal chaotic one, or will it become much more regular; how sensitive is it to change of boundary condition, of cell geometry?

At last, an other remarkable experimental result of the 1-ball experiment is that the speed $v_z$ of this ball is always larger than the piston speed $b\omega$. In our terminology, it means a "subsonic" kind of excitation. Furthermore, the ratio $v_z/(b\omega)$ is much larger in the case of resonance than within erratic motion. This is in agreement with 1-d simulations and theoretical description [20]. This shows also that the "supersonic" excitation observed when a small amount of beads fills the cell is linked to the dissipative nature of the granular gas, which is induced by ball-ball collision.

♦  *few ball dynamics:  the case of a granular gas of interacting particles in the Knudsen regime*

In a third step, the case of a granular dissipative gas has been investigated experimentally in the range of the Knudsen regime when the cell contains few balls and for different number N of balls. The distribution of ball speeds f(v) and of waiting times $P_{lid}(\tau)$ between ball-lid collisions have been determined experimentally. They both seem to exhibit an exponential law, *i.e.* $f(v) \propto \exp\{-v/v_o\}$ , $P_{lid}(\tau) \propto \exp\{-p_o\tau\}$. Both $p_o$ and $v_o$ vary linearly on the piston speed $b\omega$. Furthermore, $v_o$ is found to decrease when the number of balls N increases, while $p_o$ does not vary linearly with N; best fits using power-law have been tempted for both variables $v_o$ and $p_o$ as a function of N.

The two exponential trends, for f(v) and $P_{lid}(\tau)$, have been explained However, some incoherence in the data remains for both experimental results so that more investigations are required to suppress the remaining doubts.

So, the speed distribution f(v) is found to be not Maxwellian ; this is in agreement with what has been told in the first part, when studying the Boltzmann equation of transport. A first heuristic model has been proposed to explain the f(v)∝exp{-v/v₀}; it is based on the notion of "velostat" [7]. A second explanation has been proposed also, which is based on a direct calculation of the leading process involved; indeed, in this model balls with fast speeds are generated by an amplification due to few successive collisions with the piston without any collision with other balls; this explains the exponential trend.

This modelling has led to discuss the different definitions of cross-sections of collision, depending on the main process involved: dissipation *vs.* randomisation of speed distribution....

In conclusion, the amplitude of the typical speeds $v_0$ of the balls in this Knudsen regime has been found to vary linearly with bω and to be of the same order than the piston speed, *i.e.* $v_0 \approx b\omega$. This shows the strong effect of dissipation, which forces the working point of this system to be in the vicinity of the supersonic excitation even when ball-ball collisions are not likely. Hence this confirms the supersonic excitation observed in the MiniTexus 5 experiments for denser systems.

We conclude by telling that our understanding of the low density limit of interacting or non-interacting granular gas is still limited, and shall be improved experimentally; And the help of simulations and theory will be well come. Anyhow, one can assert that dissipation changes completely the problem of transport and communication in such random systems.

Such a result may have also some implication in the domain of theoretical economy.

*Acknowledgements:* CNES and ESA are thanked for they strong support. for funding the series of parabolic flights in board of the Airbus A300-0g. for Mini-Texus 5 rocket flight and Maxus 5 flight. The experimental results have been obtained through a cooperation involving D. Beysens. E. Falcon. S. Fauve and Y. Garrabos. with technical team assistance by C. Lecoutre and F. Palencia.

## Appendix A:

## From Liouville equation to Boltzmann equation [12, 13]

### *A.1 The case of point-like particles without interaction*

In this appendix the problem of the dynamics of a system of N interacting particles in a vibrating container is settled starting from the Liouville's equation. At the first stage no interaction is considered and the particles are defined by their position $q_i$ and momentum $p_i$, *i.e.* 6 coordinates. Each particle obeys the equation of dynamics, *i.e.* the equations of Hamilton, stating that $\partial q_i/\partial t = \partial H/\partial p_i$ and $\partial p_i/\partial t = -\partial H/\partial p_i$. So. one asks what is the typical dynamics of this system. This can be answered using the formalism of Liouville's equation, in which the system is described through its distribution function $F(\ldots q_i.p_i\ldots t)$ . So, F depends on all the positions $q_i$ and momentums $p_i$ of the spheres; it gives the probability $F \Pi dq_i dp_i$ of finding the system in a small volume of the phase space $\Pi dq_i dp_i$ at time t. Owing to the fact that the container is vibrating, the boundary condition depends upon time. One has the normalisation:

$$\int F \, \Pi dq_i dp_i = 1 \qquad (A.1)$$

$\Pi dq_i dp_i$ will be noted dqdp here after. The number of particles in a small volume $\varpi$ of the container is given by:

$$N \int_{\varpi} F(q,p,t) \, dqdp = n \qquad (A.2)$$

The variation of this number n of particles in v with time is given by

$$dn/dt = N \int_{\varpi} \partial F/\partial t \, dqdp \qquad (A.3.a)$$

But this evolution corresponds neither to a destruction nor to a creation of particles, so that it corresponds to a flux through the boundary of v. Be S the surface of the volume $\varpi$. One can write then:

$$dn/dt = -N \int_{S} F \, \mathbf{u.n} \, dS \qquad (A.3.b)$$

where $\mathbf{u}$ is the vector with 6N coordinates $\{\ldots \partial q_i/\partial t, \partial p_i/\partial t \ldots \}$ and $\mathbf{n}$ is the vector normal to the surface S, which is defined also by 6N coordinates. In Eq. (A.3.b) the surface integral can be changed in a volume integral. One gets:

$$dn/dt = -N \int_{\varpi} \nabla(F \, .\mathbf{u}) \, dqdp \qquad (A.3.c)$$

Equating Eqs. (A.3.a) and (A.3.c). and reminding that this shall be valid whatever the volume $\varpi$ be, one gets the local conservation equation of flux:

$$\partial F/\partial t + \nabla(F \, .\mathbf{u}) = 0 \qquad (A.4)$$

The operator  has 6N coordinates. So one can write

$$\nabla(F.\mathbf{u}) = \sum \partial[F(\partial q_i/\partial t)]/\partial q_i + \sum \partial[F(\partial p_i/\partial t)]/\partial p_i \qquad (A.5.a)$$

$$\nabla(F.\mathbf{u}) = \sum \{(\partial F/\partial q_i)(\partial q_i/\partial t) + (\partial F/\partial p_i)(\partial p_i/\partial t)\} + \sum F \{\partial(\partial q_i/\partial t)/\partial q_i + \partial(\partial p_i/\partial t)/\partial p_i\}$$

Last sum is 0, owing to the equations of Hamilton $\{\partial q_i/\partial t = \partial H/\partial p_i.\ \partial p_i/\partial t = -\partial H/\partial q_i\}$. This leads to:

$$\nabla(F.\mathbf{u}) = \sum \left[(\partial F/\partial q_i)(\partial q_i/\partial t) + (\partial F/\partial p_i)(\partial p_i/\partial t)\right] \qquad (A.5.b)$$

which can be written making use of Hamilton equation:

$$\nabla(F.\mathbf{u}) = \sum \left[(\partial F/\partial q_i)(\partial H/\partial p_i) - (\partial F/\partial p_i)(\partial H/\partial q_i)\right] \qquad (A.5.c)$$

This leads to the Liouville's conservation equation for F:

$$\partial F/\partial t + \sum \left[(\partial F/\partial q_i)(\partial H/\partial p_i) - (\partial F/\partial p_i)(\partial H/\partial q_i)\right] = 0 = \partial F/\partial t + \{H.F\} \qquad (A.6.a)$$

The second term is also often written as $\{H.F\}$, where $\{\}$ are called Poisson brackets. Liouville's equation can be written in different equivalent forms, including the ones in Eqs. (4), (6.a) or (6.b), which is just obtained from developing Eq. (4):

$$\partial F/\partial t + (\nabla F).\mathbf{u} + F(\nabla.\mathbf{u}) = 0 \qquad (A.6.b)$$

In these equations it is worth recalling that the operator $\nabla$ and the vector $\mathbf{u}$ have 6N coordinates each. Owing to the equation of Hamilton $(\nabla.\mathbf{u}) = 0$; hence the Liouville's equation writes:

$$\partial F/\partial t + (\nabla F).\mathbf{u} = 0 = dF/dt \qquad (A.6.c)$$

It is worth to develop Eq. (A.6.c); u is a 6N coordinate, 3N correspond to positions $\{q_i\}$ and 3N to momentum $\{p_i\}$; so half of the $u_i$'s correspond to $u_i = \partial q_i/\partial t = v_i$, which is the particle speed, and half of the $u_i$'s correspond to $u' = \partial p_i/\partial t = X_i$, which is the force acting on particle i. So one gets from Eq. (6.d):

$$\partial F/\partial t + \sum (\nabla_{\mathbf{r_i}}F).\mathbf{v_i} + \sum X_i\ (\nabla_{\mathbf{p_i}}F) = 0 \qquad (A.6.e)$$

♦ **Boltzmann's equation: the 1 particle distribution function :** Integrating F over 6(N-1) degrees of freedom corresponding to N-1 particles, allows to find the distribution function of 1 particle.

$$\partial f/\partial t + (\nabla_{\mathbf{q}}f).\mathbf{v} + X\ (\nabla_p\mathbf{f}) = 0 \qquad (A.7)$$

where f is now the 1-particle distribution function.

♦ **Box motion:** The box itself is a moving part of the total system; hence its motion can be included in the description with the Liouville's equation. As it moves periodically, its motion can be modelled as an oscillator obeying the Hamilton equation:

$$\tfrac{1}{2}\ [kq_B{}^2 + p_B{}^2/M_B] = E_B \qquad (A.8)$$

In this case, the evolution of the complete system, *i.e.* N particles plus the box. does no more depend explicitly on time. But the number of independent parameters remains the same 6N+1, *i.e.* 6N for the particles +1 for the box, since $q_B$ and $p_B$ are related because the box works at constant energy $E_B$.

I note also that the mass $M_B$ of the box shall be much larger than the mass m of the particles in order to be allowed to neglect the variation of the box motion during the collision with the particles. This imposes also $E_B \gg \Sigma_i p_i^2/(2m)$.

## A.2 Interaction between particles

Pair interactions between particles lead to an other contribution for f (and F). It comes from the term $(\partial H/\partial q_i)$ of Eq. (A.6.a), which contains all the forces acting on particle i.

It contains then interactions between the particles ij; it can be written as $T_i + \Sigma_{ij} X_{ij}$ where T is a force which derives from a potential and where $X_{ij}$ corresponds to interaction forces between the particles.

So the integration over the N-1 particles include now a term of interactionAs the particles are independent and non interacting, the distribution F can be found from the distribution of a single particle $f_i(q_i, p_i, t)$ :

$$\partial f/\partial t + (\nabla_{q_1} f).v_1 + X_1 .(\nabla_{p_1} f_1) + \int X_{12} .\nabla_{p_1} f^{(2)} \, dp_2 dq_2 = 0 \qquad (A.7)$$

Some care has to be taken to take account of the indiscernability of the particles. (see p.84-87 of [13])

## A.3 including rotations of particles

The introduction of the rotation degree of freedom is straight forward, as long as collisions are not included. One has just to introduce 4N or 6N degrees of freedom that corresponds to the rotation $\theta_i$ and rotation speeds $\Omega_i = \partial \theta_i/\partial t$ of each particles. In the formalism of Hamilton this leads to new equations $\{\partial \theta_i/\partial t = \partial H/\partial P_i. \ \partial P_i/\partial t = -\partial H/\partial \theta_i\}$.

The phase space becomes then a bit larger; but the evolution of the distribution function remains of the same kind as in Eqs. (A.6), *i.e.* controlled by the Liouville's equation in which one shall include the $\theta_i$ and the $P_i$ in the $q_i$ and in the $p_i$ respectively.

## A.4 Including collisions with the walls: Application to the statistical mechanics of independent non interacting particles in a vibrating container

The problem of the mechanics of N particles which are contained in a single box. which are not interacting with one another, but which are in interaction with the walls is equivalent to the superposition of N times the mechanics of a single particle in a box, because the motion of any particle is independent of the motion of the others and depends only on the wall motion. This means that one can limit the study to the study of a single particle in the box.

To determine correctly this dynamics, the collision rules with the walls of the container have to be introduced correctly. Assuming that the dynamics of the collision is determinist imposes that the momentums $p_i^+$ after the collisions can be deduced from the momentums $p_i^-$ before the collision, with some internal parameter such as (i) the wall speed, (ii) the orientation $\sigma$ of the surface during the collisions, and/or (iii) some other parameter $\chi$. (For instance, $\chi$ can describe some roughness of the wall of the container). So one can write:

$p_i^+ = Q_{+-}(n,t,\chi)\, p_i^-$

As $p_i$ is a vector. $Q_{+-}$ is a matrix. This can be rewritten in the centre of mass using the speeds $v_i = p_i/m$ and $v_B = p_B/M_B$, with $M_B \gg m$:

$v_i^+ - v_B(t) = Q'_{+-}(n,\chi)\, (v_i^- - v_B)$

A simple modelling is proposed for $Q_{+-}$ and $Q'_{+-}$ in appendix B.

---

## Appendix B:

## Effect of friction and rotation on ball-plate collisions

### B.1. - 3-d modelling of the ball-plate collision

The classical theory of impact, called stereo-mechanics [24], is based on impulse momentum law for rigid bodies. So, collision characteristics can be deduced from the motion of the particles before and after the collision. The present section is limited to the collisions which occur between a plane of an infinite mass and a sphere of radius R, mass $m=4\pi\rho R^3/3$ and momentum of inertia $I=2/5\, mR^2$. Be $v^\pm$ and $\varpi^\pm$ the translation and rotation speeds of the sphere before ($v^-$, $\varpi^-$) and after ($v^+$, $\varpi^+$) the collision, in the wall frame. Let us also assume that the action of collision reduces to the one of a contact force **F** at the contact point; this means in particular that (i) the rotation of the sphere in a direction perpendicular to the surface does not generate any torque, that (ii) the force generated by the impact is localised on a single point, and not in a volume or a surface. In this case integrating the equation of dynamics over the collision duration leads to:

$$m\,(\mathbf{v}^+ - \mathbf{v}^-) = \int_{\tau_{coll}} \mathbf{F}\, dt \qquad\qquad (B.1.a)$$

$$I\,(\varpi^+ - \varpi^-) = \int_{\tau_{coll}} R\mathbf{F}\, dt \qquad\qquad (B.1.b)$$

Be $F_n$ and $F_t$ the two components of **F**, normal and tangent to the contact surface; be $v_n$ and $v_t$ the normal and tangent components of the speed of the centre of mass of the sphere. In the direction normal to contact, no torque is produced, so that the system looks like a 1-d collision. This allows to introduce the restitution coefficient $\varepsilon$ as:

$$v_n^+ = - \varepsilon \, v_n^- \qquad\qquad (B.2.a)$$

Combining Eq. (B1.a) with Eq. (B.2.a) leads to:

$$m \, (1+ \varepsilon) \, v_n^- = - \int_{\tau_{coll}} F_n \, dt \qquad\qquad (B.2.b)$$

One can model now the contact as a contact with solid friction, defined by a coefficient $\kappa$ of solid friction. In this case $F_t = \kappa \, F_n$ when the contact is sliding, and $F_t$ is oriented in the direction opposite to the speed of sliding. If the contact is not sliding, $|F_t| < \kappa \, F_n$. The speed of sliding of the contact point is $v_t - \varpi R$. So, if $v_t^- - \varpi^- R = 0$, no friction develops during the collision so that $F_t = 0$ during the collision and $v_t^+ = v_t^-$, $v^+ = v^-$. On the other hand if $v_t^- - \varpi^- R \neq 0$, $F_t$ is mobilised, at least during some part of the collision duration, and two cases exist depending whether the sliding persists during the whole collision or not. If it persists, then Eq. (B.1.a) combined with the sliding condition gives:

$$\int_{\tau_{coll}} F_t \, dt = \kappa \int_{\tau_{coll}} F_n \, dt = -m \, (v_t^+ - v_t^-) = m \, \kappa (1+\varepsilon) \, v_n^-$$

and Eq. (B.1.b) writes

$$(2/5) m R^2 \, (\varpi^+ - \varpi^-) = m R \kappa (1+\varepsilon) v_n^- \qquad \text{under continuous sliding} \qquad (B.3.a)$$

and the condition of continuous sliding imposes also:

$$(v_t^+ - R \varpi^+)/(v_t^- - R \varpi^-) > 0 \qquad\qquad (B.3.b)$$

At smaller tangential speed, sliding stops at some stage during the collision, fixing the translation speed $v_t^+$ to the rotation speed $\varpi^+$, *i.e.* $v_t^+ = R \, \varpi^+$. One can get a second relation between $v_t^+$ and $\varpi^+$ from Eq. (B.1.a) and (B.1.b): $m(v^+ - v^-)R = I \, (\varpi^+ - \varpi^-)$, so that the problem is solved; but the solution depends of the initial condition. An example is given few paragraphs later).

I now consider repeated bouncing with alternatively opposite walls to show that it shall stabilise quasi-periodic linear motion. This case is quite different from the case of a ball bouncing on the bottom plate which vibrates vertically and in 1-g, because the same condition occurs at each bouncing, and the horizontal speed and the rotation speed keeps on related by the same relation all along the bouncing, with $v_t = R\varpi$. This is not the case in alternate bouncing.

## B.2.- *Alternate bouncing: Stabilisation of the 1-d trajectory due to alternate ball-lid ball-piston collisions*

So, consider the case of the same single sphere in a vibrating parallelepiped container submitted to translation vibration (or excited by a vibrating piston); but consider the case when one observes some quasi-periodic motion of the ball, which performs then alternatively collisions with the opposite walls. In this case, friction with the walls imposes rapidly the non-sliding condition at the end of collision: so, at each wall, friction fixes $v_t^+(n)=\pm R\varpi^+(n)$, where n labels the bounce. In this equation the sign + in front of R corresponds to bouncing with the bottom wall while the sign - corresponds to bouncing with the upper wall. Now, as $v_t^+(n) = v_t^-(n+1)$ in 0-g, the alternate condition on rotation after each bouncing generates a strong dissipation and one gets rapidly $v_t^+= 0$.

As a matter of fact, the non sliding conditions impose $v_t^- =\varpi^- R$ and $v_t^+= -\varpi_t^+ R$, while percussion imposes

$2mR^2(\varpi^+ -\varpi^-)/5=R\int_{\tau_{coll}} F_t \, dt$                     and                     (B.4.a)

$m\,(v_t^+ -v_t^-)=\int_{\tau_{coll}} F_t \, dt,$                                                  (B.4.b)

since $I=2mR^2/5$. This leads to the recurrence relation :

$\varpi^- = -3\varpi^+/7$ or $v_t^+(n+1)= -3v_t^+(n)/7$

which ensures the convergence within 1 or 2 bounces since one gets after a while:

$v_t(n) =(3/7)^{n-k}\, v_{tk} = v_{tk} \exp[-0.85\,(n-k)]$                         (B.5.a)

$\varpi_t(n) =(3/7)^{n-k}\, \varpi_{tk} = \varpi_{tk} \exp[-0.85\,(n-k)]$                    (B.5.b)

### B.2.a- *Quasi horizontal motion in a box vibrated vertically*

So, due to the strong dissipation which is generated by sliding during bouncing with alternate walls, the transients with rotation is strongly dissipated. This generates a strong attraction in some direction of the phase space toward an attractive subspace, that explains the reduction of the dimension of the space. This occurs because the Lyapounov exponent associated with this degree of freedom is largely negative.

It may be interesting to look at the quasi periodic transient in 1-g: Take for instance a vibrating box (size L, frequency f and amplitude of vibration b, b<<L) which vibrates vertically and throw into it a ball with an initial speed $V_o$ mainly oriented horizontally, but with some vertical component; impose $V_o>> 2Lf$ ; then after few bouncing the trajectory of the ball is stabilised along a pseudo-periodic trajectory which is horizontal; however, the trajectory moves randomly in the vertical direction due to the collision with the vertical walls which imposes some vertical speed $v_z$<<V , with $\left| v_z \right| <2\pi fa$ , depending of the phase of the vibration at each collision.

However after a while, due to what follows: the time of flight between two bouncing increases with the number n of collisions because the normal speed $V_n$ , *i.e.* the horizontal speed, decreases with n according to $V_n=\varepsilon^n V_1$. So, under 1-g, this transient quasi-periodic behaviour stops when $V_n$ becomes of the order of $(gL)^{1/2}$. Then a second kind of periodic orbit can occur if b/L is large enough to reach periodic vertical bouncing. This one becomes stable and it is the final attractor ; so the horizontal bouncing is a transient step.

### *B.2.b- Vertical drift and Horizontal motion in a box vibrated horizontally under 1-g:*

In the same way, let us imagine a box vibrating horizontally with a ball bouncing periodically and horizontally in between two vertical walls with an approximate speed $V_h=2Lf$. Due to gravity the particle falls during free flight and changes of speed so that $\delta v_z = g/(2f)$. Applying the normal collision rules, one finds that a stationary states occurs with a typical vertical speed after the bouncing $v_z^+=R\varpi^+=5g/(8f)$, while it is $v_z^-=5g/(8f)+g/(2f) = 9g/(8f)$ just before the next collision. Performing an average upon time, one gets a mean vertical drift, with a mean velocity $\underline{V_z}=7g/(8f)$.

Hence the system looks as sedimenting at constant speed, and as being subject to a viscous drag force proportional to g.

---

### **Appendix C:**

### **1-d periodic motion: stability analysis**

### *C.1- 1-d collision rules:*

A simplified way to handle the 1-d collision problem of a sphere with a plane is to work in the barycentric frame and to introduce a restitution coefficient $\varepsilon = -v_{out}/v_{in}$ [24]. As the plane is considered to have an infinite mass the piston is the centre of mass of the system.

In the case of a sphere of radius $R_{ball}$, $d=2R_{ball}$, moving in a container closed on top by a gauge and on bottom by a vibrating piston, one can determine the speed from the knowledge of the times series $\{\ldots,t_n,\ldots\}$ of the impacts. Be $z=b_p\cos(\omega t)$ the position of the piston, with $\omega=2\pi f$ ; be $L+2R_{ball}$ the position of the gauge; one can label the collisions with the vibrating bottom by even number and with the gauge by odd numbers.

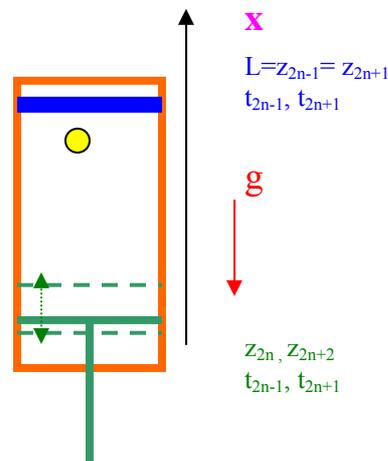

$x$
$L=z_{2n-1}= z_{2n+1}$
$t_{2n-1}, t_{2n+1}$

$g$

$z_{2n}, z_{2n+2}$
$t_{2n-1}, t_{2n+1}$

**Figure 5:**

Integration of the equation of motion in gravity field leads to $z=-\frac{1}{2}gt^2+v_ot+z_o$ one can calculate the different speeds $v_n$, $v'_n$ just before and just after each impact. One finds the relations, for the gauge:

$$v_{2n+1}= [z_{2n}-L]/(t_{2n}-t_{2n+1}) - \tfrac{1}{2}\, g(t_{2n}-t_{2n+1}) \qquad (C.1)$$

$$v'_{2n+1}= [z_{2n+2}-L]/(t_{2n+2}-t_{2n+1}) - \tfrac{1}{2}\, g(t_{2n+2}-t_{2n+1}) \qquad (C.2)$$

and for the piston:

$$v_{2n} = [L-z_{2n}]/(t_{2n-1}-t_{2n}) - \tfrac{1}{2} g (t_{2n-1}-t_{2n}) \qquad (C.3)$$

$$v'_{2n} = [L-z_{2n}]/(t_{2n+1}-t_{2n}) - \tfrac{1}{2} g (t_{2n+1}-t_{2n}) \qquad (C.4)$$

with $z_{2n} = b_p \cos(\omega t_{2n})$, and where g is negative in the Figure. One finds then for the restitution coefficients:

$$\varepsilon_{gauge} = - v'_{2n+1}/v_{2n+1} \qquad (C.4.a)$$

$$\varepsilon_{piston} = - [v'_{2n} + b_p \omega \sin(\omega t_{2n})]/[v_{2n+1} + b_p \omega \sin(\omega t_{2n})] \qquad (C.4.b)$$

## C. 2- *Stability of the periodic motion.*

In this sub-section, for sake of simplicity, the investigation of the stability is limited to the case of a 1-d periodic motion using a zero-g modelling with two different restitution coefficients $\varepsilon_p$ and $\varepsilon_g$ for the impact with the piston ($\varepsilon_p$) and the gauge ($\varepsilon_g$). For simplicity, the notation of the piston phase is changed by $\pi$ and the notation about the different speeds are changed. One writes the coordinate of the piston as $z = - b_p \cos(\omega t)$.

Be $v_n$ the speed just after the collision numbered n, with the piston; it occurs at time $t_n$, $t_{n+1}$ at coordinates $z_n = - b_p \cos(\omega t_n)$.

Be $u_n$ the speed just after the collision with the gauge; one has $u_n = - \varepsilon_g v_n$ ; and after a roundtrip the speed has become :

$$v_{n+1} - b_p \omega \sin(\omega t_{n+1}) = - \varepsilon_p [-\varepsilon_g v_n - b_p \omega \sin(\omega t_{n+1})]$$

and the time satisfies:

$$[L + b_p \cos(\omega t_n)]/v_n + [L + b_p \text{co}(\omega t_{n+1})]/(\varepsilon_g v_n) = t_{n+1}-t_n$$

So the dynamics obeys the recurrence equations:

$$v_{n+1} = \varepsilon_p \varepsilon_g v_n + (1+\varepsilon_p) b_p \omega \sin(\omega t_{n+1}) \qquad (C.6.a)$$

$$L(\varepsilon_g+1) + b_p \varepsilon_g \cos(\omega t_n) + b_p \cos(\omega t_{n+1}) = \varepsilon_g v_n (t_{n+1}-t_n) \qquad (C.6.b)$$

Exact periodic motion requires $v_{n+1}=v_n=v$ and $(t_{n+1}-t_n)=T$, where $T=2\pi/\omega$ is the period of excitation; writing $\omega t_n = \omega t_{n+1} = \chi$, one obtains:

$$v = b_p \omega (1+\varepsilon_p) \sin(\chi)/[1- \varepsilon_p \varepsilon_g]$$

$$v = b_p \omega(1+\varepsilon_g) [L/b_p +\cos(\chi)]/(2\pi \varepsilon_g)$$

In practice, $\varepsilon_p$ and $\varepsilon_g$ are not far from 1 and $L/b_p$ is large, *i.e.* >>1; so $v \approx (1+\varepsilon_g)L/(T \varepsilon_g)$ and the periodic solution is such that $\chi$ obeys:

♦ *periodic solution:*

$$2\pi \; \varepsilon_g \, (1+\varepsilon_p) \; \sin(\chi) = (1+\varepsilon_g) \, [1-\varepsilon_p \, \varepsilon_g] \, [L/b_p +\cos(\chi)] \qquad \text{(C.7.a)}$$

$$v = \; b_p \, \omega \, (1+\varepsilon_p) \sin(\chi)/[1-\varepsilon_p \, \varepsilon_g] = b_p \, \omega(1+\varepsilon_g) \, [L/b_p +\cos(\chi)]/(2\pi \; \varepsilon_g) \qquad \text{(C.7.b)}$$

As $\sin(\chi)<1$, one gets the approximate threshold condition:

♦ *Threshold condition for periodic motion:*

$$b_p/L > (1+\varepsilon_g) \, [1-\varepsilon_p \, \varepsilon_g]/[\, 2\pi \; \varepsilon_g \, (1+\varepsilon_p)] \qquad \text{(C.8)}$$

♦ *Effect of perturbation on periodic motion: stability analysis*

One can also investigate the stability of the periodic motion by linearising Eqs. (C.6) around the periodic solution and writing

| $\chi_n = \chi + \delta\chi_n$ | $v_n = v + \delta v_n$ | $\chi_n = \omega t_n$ (mode $2\pi$) | (C.9) |
|---|---|---|---|

This leads to:

$$\delta v_{n+1} = \varepsilon_p \, \varepsilon_g \, \delta v_n \; + (1+\varepsilon_p) \, b_p \, \omega^2 \delta t_{n+1} \cos(\chi) \qquad \text{(C.10.a)}$$

$$- b_p \, \varepsilon_g \; \omega\delta t_n \sin(\chi) - b_p \, \omega\delta t_{n+1} \sin(\chi) = \varepsilon_g v \, (\delta t_{n+1} - \delta t_n) \qquad \text{(C.10.b)}$$

which leads to

$$\delta v_{n+1} - (1+\varepsilon_p) \, b_p \, \omega \cos(\chi) \, \delta\chi_{n+1} \; = \varepsilon_p \, \varepsilon_g \, \delta v_n$$

$$[b\omega \sin(\chi) + \varepsilon_g \, v \,]\delta\chi_{n+1} = \; [v - b_p \, \omega \; \sin(\chi) \,] \, \varepsilon_g \, \delta\chi_n$$

and to

$$\delta v_{n+1} = \varepsilon_p \, \varepsilon_g \, \delta v_n + (1+\varepsilon_p) \, b_p \, \omega \cos(\chi)\varepsilon_g \, [v - b_p \, \omega \; \sin(\chi) \,] /[\, \varepsilon_g \, v + b_p \, \omega \sin(\chi) \,]\delta\chi_n$$

$$\delta\chi_{n+1} = \; \delta\chi_n \; \varepsilon_g \, [v - b_p \, \omega \; \sin(\chi) \,] /[\, \varepsilon_g \, v + b_p \, \omega \sin(\chi) \,]$$

So, one can write the collision rules in a matrix form:

$$\delta v_{n+1} = \alpha \; \delta v_n \; + \beta \; \delta\chi_n \qquad \text{(C.11.a)}$$

$$\delta\chi_{n+1} = \; \gamma \; \delta\chi_n \qquad \text{(C.11.b)}$$

with the following coefficients

$$\alpha = \varepsilon_p \, \varepsilon_g \qquad \text{(C.11.c)}$$

$$\gamma = \varepsilon_g \, [v - b_p \, \omega \; \sin(\chi) \,] /[\, \varepsilon_g \, v + b_p \, \omega \sin(\chi) \,] \qquad \text{(C.11.d)}$$

$$\beta = (1 + \varepsilon_p)\, b_p\, \omega\, \cos(\chi)\varepsilon_g\, [v - b_p\, \omega\, \sin(\chi)]\, /[\, \varepsilon_g\, v + b_p\, \omega\, \sin(\chi)]\qquad\text{(C.11.e)}$$

Here $\alpha$ is always smaller than 1 . $\gamma$ is smaller than 1 if the phase is in the range $0 < \chi < \pi/2$, with the convention of the vibration motion being $z = -\, b_p\, \cos(\omega t)$ and the gauge position being at $z = L$. So, the series converges absolutely in principle. However, $\gamma$ is not far from unity; therefore noise can destabilise the periodic motion in some case. This can be the case especially when $\varepsilon_g \approx 1$ and $L \gg 2\pi b_p$ .

## C.3- 1-d simulation:

Fig. 11, which is Fig.6 in [20], reports results of 1-d simulation of the dynamics of a single bead in a vibrating box under zero-g gravity, with the same restitution coefficient $\varepsilon = v_2/v_1$ at each end of the box. The calculation does not consider any viscous drag; collisions are successively computed from the knowledge of the speed of the sphere just after the previous collision and its time of occurrence, just by solving the equation of dynamics; the calculation assumes also some restitution coefficient to compute the speed after the next collision. This is a fast algorithm which is fast when the number of ball is small.

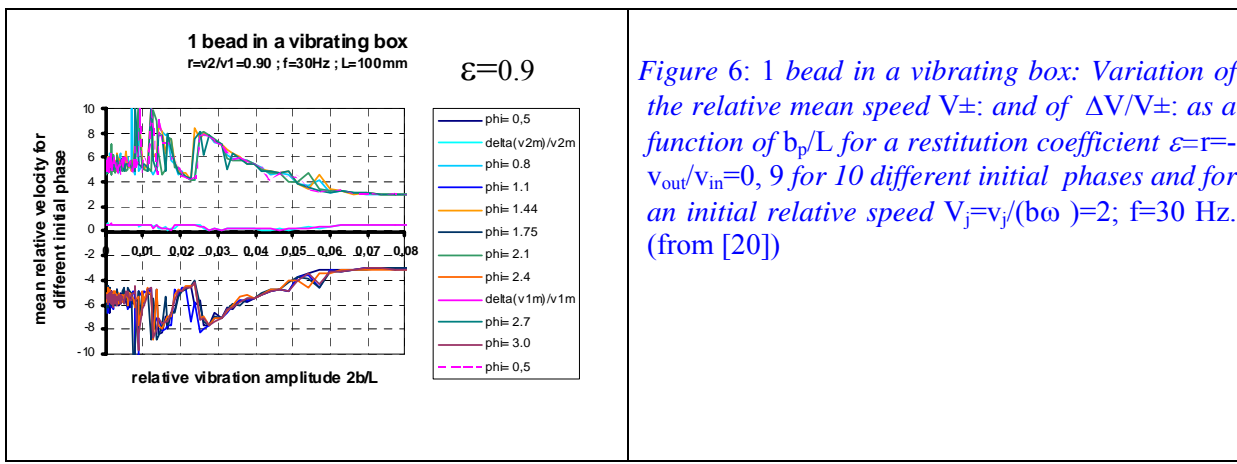

*Figure* 6: *1 bead in a vibrating box: Variation of the relative mean speed V±: and of ΔV/V±: as a function of $b_p/L$ for a restitution coefficient $\varepsilon = r = -v_{out}/v_{in} = 0, 9$ for 10 different initial phases and for an initial relative speed $V_j = v_j/(b\omega) = 2$; f=30 Hz. (from [20])*

Fig. 11 gives the mean speed v of the ball and its typical width distribution $\Delta v$ [20] in reduced unit $V = v/(2\pi a f)$ and $\Delta V = <V^2> - <V>^2$. It shows that some burst of resonance can exist. In fact, these resonances correspond to a periodic motion of the particle, which hits the piston periodically at the period of vibration ($T = 1/f$), with a phase which depends on the vibration amplitude and on the restitution coefficient. In the limit of small amplitude $b_p$, this resonance fixes the speed v of the ball to be approximately $v = 2\, f(L-d)$.

The stability of the 1-d periodic motion can also be studied analytically, as it is performed below.

## Appendix D:

## Elastic collision rules in frontal collisions

### D.2 Recall on General theory of Elastic collision:

Let us consider the elastic frontal collision of a sphere of radius R ( mass $m=4\pi\rho R^3/3$) with a very massive plane, *i.e.* M>>m. The sphere is supposed to be in relative translation motion only (*i.e.* no rotation) and to hit the plane with an initial relative speed v perpendicular to the plane (no speed component parallel to the plane). It is also assumed that v is much smaller than the sound speed c in the plate and/or in the sphere, so that the collision occurs in quasi static condition.

In this case, the collision generates the progressive deformation of the plane and of the sphere, so that (i) an area of contact is generated, which has a circular shape due to the symmetry of the problem, with a radius a, and so that (ii) the distance $R-h_1$ from the sphere centre to the plane becomes slightly smaller than R, with $2h_1 \approx a^2/R$ in the limit of small deformation. Similar compression $h_2$ occurs for the plane. Hertz's theory of elastic contact relates the compression $h_1+h_2=h$ of the two surfaces to the force F between them. According to this theory, one obtains the distribution of stress $\sigma_{xx}$ in the normal direction and the total force F in the direction normal to contact, a calculation first performed by Hertz [24-27]. Be $E_1$ and $E_2$ the Young's modulus of each material and $\nu_1$, $\nu_2$ their Poisson's ratio and defining the reduced modulus $E_r$ as $1/E_r = 3/4$ {(1-$\nu_1^2$)/($E_1$)+ (1-$\nu_2^2$)/($E_2$)}, one gets:

$$a= F^{1/3} E_r^{-1/3} R^{1/3} \qquad (D.1.a)$$

$$h= F^{2/3} E_r^{-2/3} R^{-1/3}=a^2/R \qquad (D.1.b)$$

and a maximum pressure $p_{xo}$ at the centre of the contact zone:

$$p_{xo}=3F/(2\pi a^2) \qquad (D.1.c)$$

This can be used in combination with the equation of dynamics $F=m\ d^2h/dt^2$ to solve the time dependence of the contact; it leads to define an elastic potential U which correspond to the contact interaction.

$$U=(2/5)\ h^{5/2}\ E_r\ R^{1/2} \qquad (D.2.a)$$

Defining the parameter $k= 4RE_r/5$, one finds also the characteristic time $\tau_{coll}$ of the collision and the maximum deformation $h_{max}$ and area of contact $S_{max}=\pi a_{max}^2= \pi h_{max}R$.

$$\tau_{coll} \approx 2.94\ m^{2/5}\ k^{-2/5}\ v^{-1/5} \qquad (D.2.b)$$

$$\tfrac{1}{2}\ mv^2=(2/5)\ h_{max}^{5/2}\ E_r\ R^{1/2} \qquad (D.2.c)$$

One infers the scaling relations from Eqs. (D.2.c) and (D.2.b):

| $S_{max} = \pi a_{max}^2 \approx \pi R h_{max} \approx R^2 v^{4/5}$ | $\tau_{coll} \approx m^{2/5} v^{-1/5}$ | (D.2.d) |
|---|---|---|

When the speed of the particle is large, the limit of validity of elasticity may be overpassed and plastic deformation can be generated. This occurs when the stress field is larger than a limit $\sigma_p$. Imposing $p_{xo} < \sigma_p$ in Eq. (D.1.c) and as $F = E_r a^3/R$ according to Eq. (D.1.a), one infers the condition:

$$p_{xo} = 3E_r a_{max}/(2\pi R) = (h_{max}/R)^{1/2} [3E_r/(2\pi)] = (5/32)^{1/5} (3/\pi)^{4/5} [\rho^{1/5} E_r^{-4/5} v^{2/5}] < \sigma_p$$

## D.2  Elastic collision preserves some conservation rules:

One can find that the general following rules are preserved during elastic collisions:

$$a^5/v^2 = 1.25 \ R^2/E_r \qquad\qquad\qquad\qquad (D.3.a)$$

$$a \ \tau_{coll}^2 = 19.19 \ m/E_r \qquad\qquad\qquad\qquad (D.3.b)$$

$$a^2 /(v \ \tau_{coll}) = 0.602 \ R \qquad\qquad\qquad\qquad (D.3.c)$$

| $a = F^{1/3} E_r^{-1/3} R_1^{1/3}$ | $h = F^{2/3} E_r^{-2/3} R^{-1/3}$ | $U = (2/5) h^{5/2} E_r R^{1/2}$ | $\tau_{coll} = 2.94 \ m^{2/5} k^{-2/5} v^{-1/5}$ | $\sigma_z = 3F/(2\pi a^2)$ |
|---|---|---|---|---|
| | | $U_{max} = (2/5)(2\pi/3)^5 E_r^{-4} \sigma_p^5 R^4/\rho \approx m v_{max}^2 /\alpha$ | | |
| $E = 160$ GPa | $\nu = 0.2$ (?) | $\sigma_p = 0.3$GPa to 2GPa | $\rho = 8 \ 10^3$ kg/m$^3$ | |
| $E_r = 2E/[3(1-\nu^2)]$ $E_r = 1.11 \ 10^{11}$ Pa | $k = 4 \ E_r R/5$ $k = 888 \ 10^8 R$ (SI) | $c^2 = [E(1-\nu)] /[\rho(1+\eta)(1-2\eta)] \approx E/\rho$ $c = 4 \ 710$m/s | $m = 4\pi\rho R^3/3$ | |

**Table D.1:** Mechanical characteristics of steel and of an elastic frontal collision between a sphere and a plane: a stands for the radius of the contact area, h for the thickness of the deformed zone of contact, U for the elastic energy, t$_{coll}$ for the duration of elastic collision, $\sigma_z$ for the stress, $\sigma_p$ for the plastic stress, F for the force, $\nu$ for the Poisson coefficient of steel, E for the Young's modulus, $\rho$ for the density, R for the ball radius, c for the sound speed

In fact it seems that the last conservation relation remains valid even when dissipation occurs [28].
Table D.1 sums up the relations obtains for "Hertz frontal collision", and gives numerical estimates for stainless steel.

***Acknowledgements:*** CNES and ESA are thanked for they strong support. for funding the series of parabolic flights in board of the Airbus A300-0g. for Mini-Texus 5 rocket flight and Maxus 5 flight. The experimental results have been obtained through a cooperation involving D. Beysens. E. Falcon. S. Fauve and Y. Garrabos. with technical team assistance by C. Lecoutre and F. Palencia.

**Errata:** in Appendix C, labelling of Fig. 5 and 11 (or 6 of ref 20) is incorrect, please read Fig C1 and C2 respectively.

The electronic arXiv.org version of this paper has been settled during a stay at the Kavli Institute of Theoretical Physics of the University of California at Santa Barbara (KITP-UCSB), in june 2005, supported in part by the National Science Fundation under Grant n° PHY99-07949.

*Poudres & Grains* can be found at :
http://www.mssmat.ecp.fr/rubrique.php3?id_rubrique=402